\documentclass[12pt,superscriptaddress]{iopart}

\usepackage{cite}
\usepackage[dvips]{graphicx}
\usepackage{color}

\newcommand{\figurewidth}{0.5}

\newcommand{\bra}[1]{\langle\,{#1}\, |}
\newcommand{\ket}[1]{|\,{#1}\,\rangle}
\newcommand{\braket}[2]{\mbox{$\langle\,{#1}\, | \,{#2}\,\rangle$}}

\newcommand{\frefp}[2]{Fig.~\ref{#1}(#2)}

\newcommand{\ssection}[1]{{\noi  \it #1:}}

\newcommand{\sub}[2]{ {#1}_{\mbox{\scriptsize #2}} }

\newcommand{\bv}[1]{\mathbf{ #1 }}

\def\noi{\noindent}
\def\beq{\begin{equation}}
\def\eeq{\end{equation}}

\def\figurewidth{0.99}

\newcommand{\rref}[1]{Ref.~\cite{#1}}
\newcommand{\bref}[1]{(\ref{#1})}

\newcommand{\cref}[1]{chapter~\ref{#1}}
\newcommand{\Cref}[1]{Chapter~\ref{#1}}
\newcommand{\aref}[1]{\ref{#1}}

\begin{document}

\title[Rydberg aggregates]
{Rydberg Aggregates}

\author{S. W{\"u}ster$^{1,2,3}$ and J.-M. Rost$^1$} 
\address{$^1$Max-Planck-Institute for the Physics of Complex Systems, 01187 Dresden, Germany} 
\address{$^2$Department of Physics, Bilkent University, Ankara 06800, Turkey}
\address{$^{3}$Department of Physics, Indian Institute of Science Education and Research, Bhopal, Madhya Pradesh 462 023, India}
\ead{sebastian@iiserb.ac.in}
\begin{abstract}
We review Rydberg aggregates, assemblies of a few Rydberg atoms exhibiting energy transport through collective eigenstates, considering isolated atoms or assemblies embedded within clouds of cold ground-state atoms. We classify Rydberg aggregates, and provide an overview of their possible applications as quantum simulators for phenomena from chemical or biological physics. Our main focus is on flexible Rydberg aggregates, in which atomic motion is an essential feature. In these, simultaneous control over Rydberg-Rydberg interactions, external trapping and electronic energies, allows Born-Oppenheimer surfaces for the motion of the entire aggregate to be taylored as desired. This is illustrated with theory proposals towards the demonstration of joint motion and excitation transport, conical intersections and non-adiabatic effects. 
 Additional flexibility for quantum simulations is enabled by the use of dressed dipole-dipole interactions or the embedding of the aggregate in a cold gas or BEC environment.
Finally we provide some guidance regarding the parameter regimes that are most suitable for the realization of either static or flexible Rydberg aggregates based on Li or Rb atoms. The current status of experimental progress towards enabling Rydberg aggregates is also reviewed.
\end{abstract}
%


\section{Introduction and overview}\label{intro}
%
Ultra cold gases in which some atoms have been highly excited to electronic Rydberg states \cite{book:gallagher} are rapidly becoming an important pillar of ultra cold atomic physics. Through strong and long range dipole interactions between atoms in a Rydberg state and interaction blockades \cite{lukin:quantuminfo,urban:twoatomblock,gaetan:twoatomblock,tong:blockade,cenap:manybody,singer:blockade}, they enable, for example, studies of strongly correlated spin systems \cite{Zeiher:ryd_dress_spinchain,Glaetzle:spinice:prx}, glassy relaxation as in soft-condensed matter \cite{lesanovsky:kinetic}, many-body localization \cite{Marcuzzi:manybodyloc} or non-local and non-linear (quantum) optics \cite{Mohapatra:giantelectroopt,sevincli:nonlocopt,peyronel:quantnonlinopt,dudin:singlephotsource}. Apart from these links to quantum many-body phenomena, also the relation of cold Rydberg systems to classical many-body physics has been explored, for example through the creation of strongly coupled ultracold plasmas \cite{Killian:science,Killian:ucolplasma:PhysRevLett.83.4776,heidelberg:plasmaexp,Bannasch:strongcoup:PhysRevLett.110.253003,pohl:plasmacrystal:PhysRevLett.92.155003}.  Review articles on cold Rydberg atoms \cite{loew:rydguide:jpbreview,lim:rydbergreview}, their exposure to strong magnetic fields \cite{pohl:rydBfield:physrep}, their role in ultracold plasmas \cite{kilian:ucolplasma:review} as well as their application to quantum information \cite{Saffman:quantinfryd:review} and nonlinear quantum optics \cite{firstenberg:rydberg:NLquantoptreview} are available.

\begin{figure}[htb]
\centering
\includegraphics[width= 0.7\columnwidth]{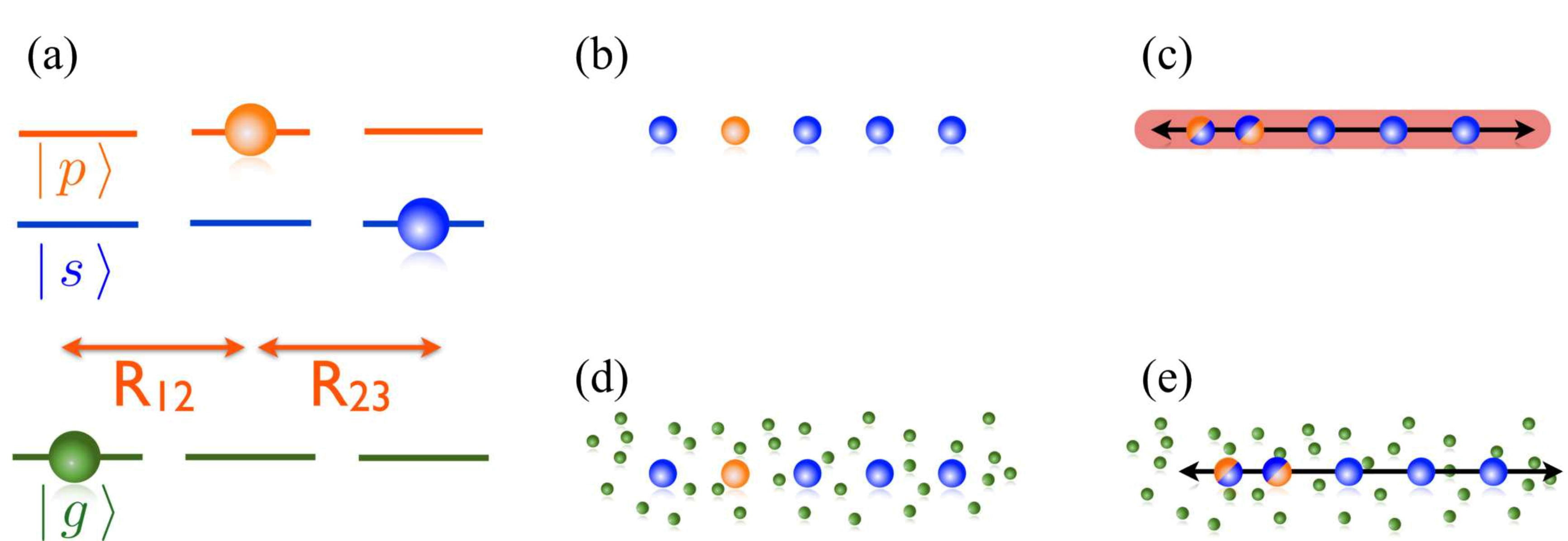} 
\caption{Rydberg aggregates and their classification. (a) Essential ingredients and level scheme, (green balls) ground state atoms, (blue balls) Rydberg $\ket{s}$ states, (orange balls) Rydberg $\ket{p}$ states. Distances $R_{ij}$ between locations will be crucial later. We classify these aggregates as follows: (b) Static Rydberg aggregate. 
(c) Flexible Rydberg aggregate with atoms free in quasi-1D, two-colored balls indicate a superposition state $\sim\ket{sp}\pm\ket{ps}$. (d)  Embedded Rydberg aggregate. (e) Embedded flexible Rydberg aggregate.
\label{overview_sketch}}
\end{figure}
This review covers a complementary subfield of Rydberg physics, namely assemblies of Rydberg atoms that share collective excitations through resonant dipole-dipole interactions, e.g.~involving angular momentum $\ket{s}$ and $\ket{p}$ states.  For reasons which will become obvious below, we have dubbed those assemblies {\em Rydberg aggregates}.
They can consist of individually trapped atoms \cite{nogrette:hologarrays,barredo:trimeragg}, or Rydberg excitations in a cold atom cloud, which acts as a host to the aggregate, e.g.~\cite{cenap:emergent,maxwell:polaritonstorage}. Both scenarios are depicted in \fref{overview_sketch}. In the latter case, the position of Rydberg atoms can still be controlled through the interplay of interaction blockade and cloud volume \cite{schauss:mesocryst}, background gas induced Rydberg energy shift \cite{middelkamp:rydinBEC,balewski:elecBEC} or using very tight excitation laser foci. The aggregate atoms are in close enough proximity to experience significant dipole-dipole interactions involving resonant states, if more than a single Rydberg angular momentum state is present in the system. 
We can then realize two regimes: (i) Static Rydberg aggregates, with excitation transport through long-range interactions occurring fast, before atoms can be set into motion. This corresponds to the frequently cited ``frozen gas regime''. (ii) Flexible Rydberg aggregates, where one deliberately waits until dipole-dipole interactions have also set the atoms themselves into motion. Both, static as well as flexible Rydberg aggregates can be realized in isolation or embedded in a cold gas cloud, leading to the four possible scenarios sketched in \frefp{overview_sketch}{b-e}.
Together, these four types of Rydberg aggregates offer intriguing opportunities to create and study coherent energy-, angular momentum- and entanglement transport as well as, on a more formal level, the formation of Born-Oppenheimer surfaces governing the motion of the Rydberg aggregate atoms, all due to the existence of collective excited states enabled through strong long-range interactions. 

Static Rydberg aggregates can be described with the help of exciton theory well known from molecular aggregates, as we will see. Flexible Rydberg aggregates
add qualitatively new features, by taking the mechanism through which stable molecules form (an intricate interplay of bound coherent nuclear and electronic motion) to coherent dynamics in the continuum.
We will review  the basic mechanism for exciton formation, exciton transport and inter-atomic forces at work in Rydberg aggregates, summaries existing experimental studies and review theory proposals on their potential uses. These range from understanding quantum transport on static networks, over transport in the presence of decohering environments to directed entanglement transport. 

\begin{figure}[htb]
\centering
\includegraphics[width= 0.7\columnwidth]{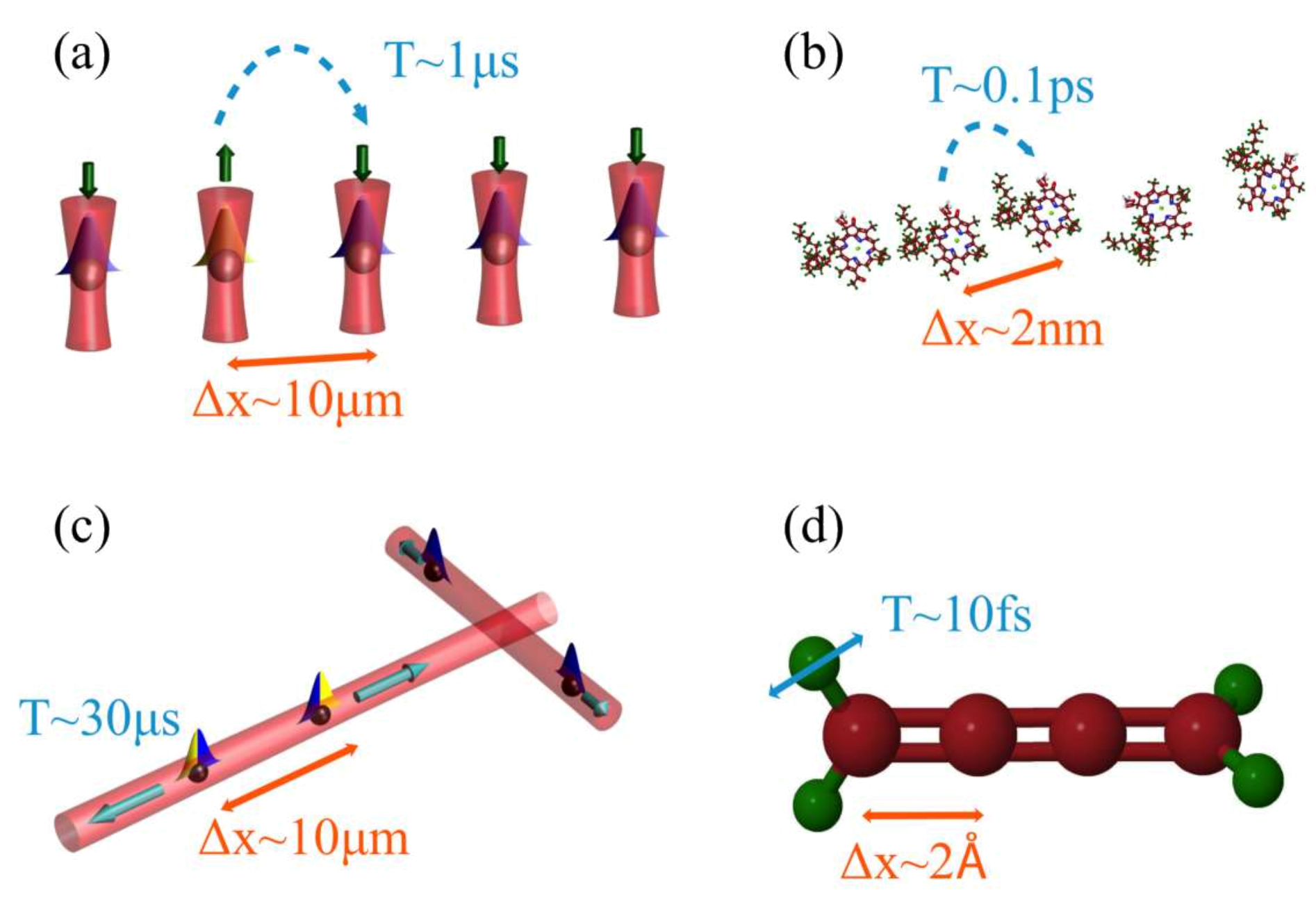} 
\caption{Rydberg aggregates in a bigger context. (a) Static atomic aggregate versus (b) molecular aggregate. The underlying physical energy transport principles are identical, but characteristic scales differ markedly. The characteristic spatial (temporal) scales of excitation migration are shown as orange (blue) text. 
 (c) Flexible Rydberg aggregates of atoms moving in some constrained geometry. We sketch two one-dimensional red detuned optical traps acting as linear chain for atom movement as red shaded cylinders.  Scales there can be compared with those in (d), of nuclear motion on the Born-Oppenheimer surfaces of a complex molecule. 
\label{motivation}}
\end{figure}

The broader vision that appears ultimately reachable in these systems is to enable a quantum simulation platform for processes in (bio)-chemistry as presented in \fref{motivation}. A static Rydberg aggregate, as sketched in panel (a), shares the fundamental energy transport mechanism with photosynthetic light harvesting complexes \cite{grondelle:review,grondelle:book} or molecular aggregates \cite{saikin:excitonreview,kirstein:Jaggregates,kuehn_lochbrunner:excitonquantdyn}, sketched in (b). However the spatial- and temporal scales of energy transport in the Rydberg atomic and molecular settings are extremely different. Moreover, the biochemical transport process is  typically severely affected by decoherence, while the process in ultracold atoms is not. However, since the introduction of controllable decoherence into the Rydberg setting is possible, the easier accessible scales of the Rydberg system should allow for detailed model studies of the interplay of transport and coherence. Similarly we can draw connections between the motion of atoms in a flexible Rydberg aggregate, shown in panel (c) and the motion of nuclei on a molecular Born-Oppenheimer surface (d). Using similar scale arguments as above, flexible Rydberg aggregates hold promise to give experimentally controlled access to the analogue of quintessential processes from nuclear dynamics in quantum chemistry.

%
%
We focus in this review article more strongly on  flexible Rydberg aggregates, see \fref{overview_sketch}(c), and their links to molecular physics, see \fref{motivation}(d): Rydberg aggregates with constituents allowed to move enable the tailored creation of Born-Oppenheimer surfaces with characteristic length-scales of $\mu$m, vastly exceeding typical intra-molecular length-scale of Angstroms. This renders paradigmatic quantum chemical effects, such as passage through a conical intersection, accessible much more directly in experiment. Ideas originating from this novel quantum simulation platform may then in turn benefit chemistry. 

The exploration of energy transport with tuneable environments, see \fref{motivation}(a,b), is only briefly commented upon here, more details can be found in \cite{schoenleber:immag}.

The eponym of Rydberg aggregates are molecular aggregates \cite{eisfeld:Jagg,saikin:excitonreview}, due to their related energy transport characteristics \cite{eisfeld:Jagg,cenap:motion,muelken:excitontransfer}. In the molecular context, aggregation implies self-assembly in solution with subsequent mutual binding. Rydberg aggregates, in contrast are typically not bound, except in special cases \cite{zoubi:VdWagg}. However, self-assembly of Rydberg aggregates in regular structures can take place in strongly driven dissipative Rydberg systems \cite{lesanovsky:nonequil_structures,gaertner:faciitationagg:theory,malossi:fullcountstat,Simonelli:excitationavalanche:JPB,valado:exp:kineticconstraints,schempp:countingstat,urvoy:vaporaggreg}, by configuring the excitation process to develop a strongly preferred nearest neighbor distance. The term ``Rydberg aggregates'' is hence also used in the literature to refer to these ordered constructs in a dissipative setting. In contrast, our definition implies collective excited states in form of a coherently shared excitation, with direct analogy for the static Rydberg aggregates to transport physics in molecular aggregates.

The majority of cold Rydberg atomic physics experiments to date has explored the so called ``frozen gas'' regime, where thermal motion of the atoms could be neglected for the sub-microsecond duration of electronic Rydberg dynamics of interest. In experiments motivated by quantum information applications \cite{jaksch:dipoleblockade,lukin:quantuminfo,urban:twoatomblock,gaetan:twoatomblock}, motion is then a residual noise source \cite{wilk:entangletwo,mueller:browaeys:gateoptimise}.  Experiments with Rydberg aggregates were also performed in this regime \cite{barredo:trimeragg,cenap:emergent,maxwell:polaritonstorage}.
Some experiments instead created unfrozen Rydberg gases, with Rydberg atoms experiencing strong acceleration, primarily to characterize interactions and ionization, e.g.~\cite{Fioretti:longrangeforces,li_gallagher:dipdipexcit,Amthor:mechVdW,amthor:autoioniz,park:dipdipionization,Zhigang:Collisions:PRA,Faoro:VdWexplosion}. Recent experiments that permit a particularly detailed real time observation of such motion will be reviewed below \cite{thaicharoen:trajectory_imaging,thaicharoen:dipolar_imaging,celistrino_teixeira:microwavespec_motion,guenter:EITexpt}. Flexible Rydberg aggregates, in which motion of atoms is fully controlled or guided, instead of randomly seeded, can now be one of the next experimental steps. 

The review is organized as follows:  Sections 2-4 will deal with isolated Rydberg aggregates, providing the
theoretical background in section 2, and detailing static and flexible aggregates in sections 3 and 4, respectively. Section 5 will introduce embedded Rydberg aggregates and section 6 describes Rydberg dressed aggregates, an important 
possibility to extend the parameter regime for realizing Rydberg aggregates. Section 7 provides 
a landscape for the existence of static and flexible Rydberg aggregates in the plane of the two most relevant parameters that can be changed: the distance of Rydberg atoms $d$ and their principal quantum number $\nu$.
The reader may want to consult this section for orientation throughout the review to identify relevant parameter regimes. We return to our quantum simulation perspective presented in the introduction in \sref{outlook} and after concluding provide an appendix with additional technical detail.

\section{Theoretical background}\label{models}

We begin with a compact summary of the formal methods used to describe Rydberg aggregates, both static and flexible.

\subsection{Basic model}\label{basicmodel}
%
Rydberg aggregates are typically  treated with an essential state model, in which each atom of the aggregate can be in either of two highly excited Rydberg states $\ket{s}=\ket{\nu,l=0}$ or $\ket{p}=\ket{\nu,l=1}$, with a principal quantum number $\nu$ usually in the range $\nu=20\dots 100$, see the  sketch  \frefp{overview_sketch}{a}. Any other pair of adjacent angular momentum values such as $\ket{p}$, $\ket{d}$ gives the same qualitative results, at the expense of less energetic separations to states not desired in the model.
Atoms in their ground-state $\ket{g}$ occur prior to excitation, or when employed as an external bath. For most of the phenomena discussed in this review, the electronic state space of an $N$-atom Rydberg aggregate can be further simplified to the single-excitation manifold spanned by $\ket{\pi_n} =\ket{sss\cdots p\cdots sss}$, where only the $n$th atom is in $\ket{p}$ and all others are in $\ket{s}$. For parameter regimes of interest here, the most important interactions are of a dipole-dipole form
\begin{equation}
\sub{\hat{H}}{el}(\bv{R})=\sum_{n\neq m}^N H_{nm}(R_{nm})  \ket{\pi_n}\bra{\pi_m}=\sum_{n\neq m}^N \frac{C_3}{|\bv{r}_n - \bv{r}_m|^3} \ket{\pi_n}\bra{\pi_m},
\label{Hdd}
\end{equation}
where $C_3$ is the dipole-dipole dispersion coefficient and $\bv{r}_n$ the position of atom $n$. With $\bv{R}=\{\bv{r}_1,\dots,\bv{r}_N\}$ we denote the collection of all atomic coordinates. 
For simplicity we  assume isotropic dipole-dipole interactions. In \sref{conical_intersections} we provide a list of references where model extensions and consequences of anisotropies are discussed.

A general electronic state for a static aggregate can be written as $\ket{\Psi(t)} = \sum_n c_n(t) \ket{\pi_n}$. When the aggregate is initialized such that the single excitation resides on a specific site $k$, with $c_n(0)=\delta_{nk}$, the Hamiltonian \bref{Hdd} will cause a coherent wave-like transport of the excitation, typically de-localizing over all atoms in the aggregate.

Without the restriction to a single excitation, one can consider each aggregate atom as a pseudo-spin $1/2$ object, and write 
\begin{equation}
\sub{\hat{H}}{el}(\bv{R})=\sum_{n\neq m}^N \frac{C_3}{|\bv{r}_n - \bv{r}_m|^3} \hat{\sigma}_n^+ \hat{\sigma}_m^-,
\label{Hddspin}
\end{equation}
where the $\hat{\sigma}_n^\pm$ are the usual Pauli raising and lowering operators.
We discuss corrections to the simple model presented here later (\sref{interactions}). 

\subsection{Exciton eigenstates and Born-Oppenheimer surfaces}\label{excitons_surfaces}
%
In the time-independent Schr\"odinger equation 
\begin{equation}
\sub{\hat{H}}{el}(\bv{R})\ket{\varphi_n(\bv{R})} = U_n(\bv{R})\ket{\varphi_n(\bv{R})} 
\label{eigensystem}
\end{equation}
for the electronic Hamiltonian \eref{Hddspin}, eigenstates and eigenenergies parametrically depend on all co-ordinates contained in $\bv{R}$.
Eigenstates can be written as superpositions  
\begin{equation}
\ket{\varphi_n(\bv{R})}=\sum_k c_{nk}(\bv{R}) \ket{\pi_k}
\label{excitons}
\end{equation}
of the so-called \emph{diabatic} basis states $\ket{\pi_k}$, and are referred to as \emph{adiabatic} electronic basis states or (Frenkel) \emph{excitons} \cite{frenkel_exciton,frenkel:footnote}.  Hence, in an exciton state the entire aggregate typically shares the p-excitation collectively, see also footnote \cite{frenkel:footnote}. As in molecular physics, we refer to the corresponding eigenenergies $U_n(\bv{R})$ as \emph{Born-Oppenheimer surfaces}. In the context of Rydberg aggregates, these concepts were first explored in \cite{cenap:motion}.

\subsection{Quantum dynamics and surface hopping}\label{dynamics}
%
When considering motion of the Rydberg aggregate quantum mechanically, we can expand the total state $\ket{\Psi(\bv{R},t)}=\sum_{n=1}^{N} \phi_{n}(\bv{R},t) \ket{\pi_n}$ and obtain the Schr{\"o}dinger equation ($\hbar=1$) \cite{wuester:CI}
\begin{equation}
i\frac{\partial}{\partial t}   \phi_{n} (\bv{R})=\sum_{m=1}^N\left[-\frac{ \nabla^2_{\bv{r}_m} }{2 M}  \phi_{n}(\bv{R}) + H_{nm}(R_{nm}) \phi_{m}(\bv{R})\right]
\label{fullSE}
\end{equation}
with mass $M$ of the atoms. Since in this equation the dipole-dipole atoms give rise to both, excitation transport between different atoms as well as forces on these atoms, we expect an intimate link between motion and transport.

For up to about 3 spatial degrees of freedom Eq.~\eref{fullSE}  can be numerically solved  straightforwardly \cite{wuester:CI,leonhardt:switch}, e.g., by using XMDS \cite{xmds:paper,xmds:docu}. For more 
degrees of freedom the direct solution of the quantum problem typically becomes too challenging, and one has to resort to quantum-classical methods. 

Before describing those, let us first consider \bref{fullSE} expressed in the adiabatic basis, such that $\ket{\Psi(\bv{R},t)}=\sum_{n=1}^{N} \tilde{\phi}_{n}(\bv{R},t) \ket{\varphi_n(\bv{R})}$. We then obtain the Schr{\"o}dinger equation in terms of Born-Oppenheimer surfaces $U_{n}(\bv R)$,
\begin{equation}
i\frac{\partial}{\partial t}   \tilde{\phi}_{n} (\bv{R})=\left( \sum_{k=1}^N\left[-\frac{ \nabla^2_{\bv{r}_k} }{2 M}\right] +   U_n(\bv{R})  \right)\tilde{\phi}_{n}(\bv{R}) + \sum_m D_{nm}(\bv{R}) \tilde{\phi}_{m}(\bv{R}),
\label{fullSE_adiab}
\end{equation}
where
\begin{equation}
D_{nm}(\bv{R})=-\frac{1}{2M}\left( \bra{\varphi_n(\bv{R})} \nabla^2 \ket{\varphi_m(\bv{R})}+ 2\bra{\varphi_n(\bv{R})} \nabla \ket{\varphi_m(\bv{R})}\cdot \nabla\right)
\label{nonadiab}
\end{equation}
are the non-adiabatic coupling terms.

The kinetic term within square brackets in \eref{fullSE_adiab} is responsible for 
motion on the excitonic Born-Oppenheimer (BO) surface $U_n(\bv{R})$. Even in ultra-cold Rydberg atomic physics, this motion is for Rydberg atoms frequently in a regime which can be treated classically, using Newton's equations \cite{Amthor:mechVdW,amthor:modellingmech,thaicharoen:trajectory_imaging,celistrino_teixeira:microwavespec_motion}. As long as the energy separations between adjacent electronic surfaces of the atoms are large compared to kinetic energies, one typically can also ignore non-adiabatic couplings.

However, excitonic BO surfaces $U_n(\bv{R})$ derived from resonant dipole-dipole interactions frequently exhibit points of degeneracies (conical intersections, see \sref{conical_intersections}), near which the energy separation criterion cannot be fulfilled. Then, one has to take into account non-adiabatic transitions  \bref{nonadiab}. They generically lead to quantum dynamics that involves coherent superpositions of multiple occupied BO surfaces $n$. A convenient class of methods from quantum chemistry, designed to tackle this problem while retaining simple classical propagation of the nuclei, is quantum-classical surface hopping \cite{drukker:surfacehopping:review}. We focus particularly on Tully's fewest switches algorithm \cite{tully:hopping,tully:hopping2,tully:hopping:veloadjust,barbatti:review_tully}, which has proven a versatile and accurate method for Rydberg aggregates \cite{wuester:cradle,moebius:cradle,moebius:bobbels,leonhardt:switch}. 

 In Tully's method, the motion of the Rydberg atoms is still modeled classically 
\begin{equation}
M\frac{\partial^2}{\partial t^2}    \bv{r}_{k}=- \nabla_{\bv{r}_{k}} U_{s(t)}(\bv{R}),
\label{newton}
\end{equation}
on a specific Born-Oppenheimer surface, which may change in time
 through random jumps from one surface to another ($n\to m$), creating a path $s(t)$  in accordance with the probability for  non-adiabatic transitions. 
Simultaneous population of multiple surfaces is  made possible by simulating the full dynamics as a stochastic distribution from different paths $s(t)$.

Along with the motion of the Rydberg aggregate atoms, we are interested in the exciton dynamics. It is described by the time-dependent electronic wave function $\ket{\Psi(t)}=\sum_n c_{n}(t) \ket{\pi_n}=\sum_n \tilde{c}_{n}(t) \ket{\varphi_n(\bv{R}(t))}$. Analogously to \bref{fullSE} and \bref{fullSE_adiab} its time-evolution can be determined in the diabatic,
\begin{equation}
i\frac{\partial}{\partial t}   c_{n}(t) =\sum_{m=1}^N H_{nm}[R_{nm}(t)] c_{m}(t), 
\label{Tully_electron}
\end{equation}
or adiabatic basis
\begin{equation}
i\frac{\partial}{\partial t}   \tilde{c}_{n}(t) =U_n[\bv{R}(t)]  \tilde{c}_{n}(t)  + \sum_{m=1}^N d_{nm}  \tilde{c}_{m}(t)
\label{Tully_electron_adiab}
\end{equation}
with 
\begin{equation}
d_{nm}=-\frac{1}{M}\bra{\varphi_n(\bv{R}(t))} \nabla \ket{\varphi_m(\bv{R}(t))}\cdot \mathbf{v}, 
\label{non_adiab_coupl_tully}
\end{equation}
where $\mathbf{v}=\partial \mathbf{R}(t)/\partial t$ is the system velocity.

In practice it is typically simpler to work with \bref{Tully_electron}, while \eref{Tully_electron_adiab} shows more clearly that the probability for a stochastic transition from surface $n$ to surface $m$ must be subject to $d_{nm}$. It is chosen such that  the fraction of trajectories on surface $m$ at time $t$ is given by $| \tilde{c}_{m}(t)|^2$. 
Further technical details about the implementation can be found in the original work \cite{tully:hopping,tully:hopping2}, the review \cite{barbatti:review_tully} and for the context of Rydberg aggregates in  \cite{leonhardt:unconstrained,leonhardt:orthogonal} or \cite{leonhardt:thesis}.

For Rydberg aggregates, solutions with Tully's algorithm have been directly compared with full quantum dynamical solutions of \bref{fullSE} \cite{wuester:cradle,leonhardt:switch}, and showed excellent agreement in the cases tested, even if conical intersections are involved \cite{leonhardt:switch}. The likely reason is that these cases do not involve trajectories with motion visiting the same spatial point twice, that is~$\bv{R}(t+\Delta t) \approx \bv{R}(t)$ does not occur.  Therefore, spatial complex phases of the atomic (nuclear) wave function $\phi_{n} (\bv{R})$, which are not represented in quantum classical methods, do not play a role.

\subsection{Rydberg trapping}\label{trapping}

Some of the physics we review below, will require the controlled guidance of the motion of Rydberg atoms, and hence trapping of Rydberg atoms. However, the main tools for 
  trapping of ultracold ground-state atoms, optical and magnetic trapping \cite{ketterle:review,book:pethik}, essentially rely on quantitative features of the electronic ground-states, and are typically not directly applicable to the trapping of atoms in a Rydberg state.
 
Nonetheless the tools can be ported. Optical trapping of Rydberg atoms was demonstrated, e.g.,~in \cite{li:lightatomentangle}, even such that the trapping frequencies for ground- and Rydberg states coincided (magic trapping) \cite{zhang:magicwavelength,Topcu:magictrap:PhysRevA.88.043407}. For the Rydberg state the trapping arises through the averaged ponderomotive energy of the electron, see also \cite{Younge:pondlatt:NJP}.

A more flexible option would be the use of earth-alkali atoms, in which one outer electron can be addressed for trapping, while the other is exploited for Rydberg physics \cite{rick:Rydberglattice,gil:latticeclock}. Trapping through static magnetic and electric fields has been considered as well, where additional features may arise due to the significant extension of the Rydberg orbit on the spatial scale where fields vary. Therefore, nucleus and valence electron of the atom
have to be treated separately \cite{mayle:rydbergtrap:dressing_in_trap,mayle:rydbergtrap:gsprobe,mayle:rydtrap_lowL}.

\section{Static Rydberg aggregates}\label{static}

We call Rydberg aggregates ``static'' if the motion of constituent atoms can be neglected. This may be the case because inertia prevents significant acceleration of atoms on the time-scale of interest
or because atoms are trapped (\sref{trapping}). For static aggregates, \bref{Hdd} provides a simple model for the transport of a single energy quantum (the $p$-excitation) through long-range hopping on a set of discrete ``sites'' (the remaining $s$-atoms). The underlying process shares features with energy transport in molecular aggregates or during photosynthetic light harvesting, as first realized in \cite{eisfeld:Jagg} and developed further in \cite{muelken:excitontransfer}.

\subsection{Exciton theory}\label{transport_theory}

\begin{figure}[htb]
\centering
\includegraphics[width= \columnwidth]{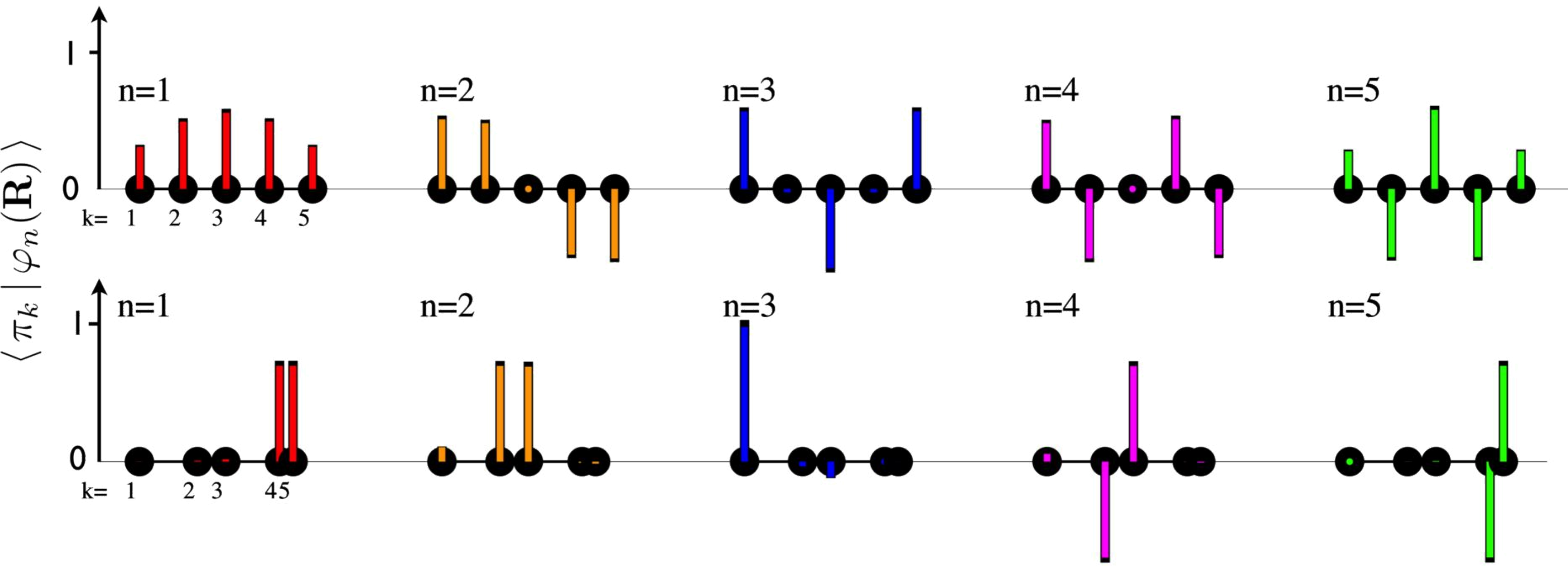} 
\caption{Frenkel excitons of a regular Rydberg aggregate with $N=5$, adapted from \cite{cenap:motion}. Bars indicate the magnitude and sign of coefficients $c_{nk}=\braket{\pi_k}{\varphi_n(\bv{R})}$ in \bref{excitons}. The colors distinguish different excitons, from left to right their energy is decreasing (assuming $C_3>0$).
\label{chain_excitons}}
\end{figure}

If atomic motion is neglected, the Hamiltonian \bref{Hdd} is time-independent and all information about the system is contained in the excitons  \bref{eigensystem} and their energies.
These states are shown in \fref{chain_excitons} (top) for the example of a chain of five Rydberg atoms with equidistant spacing $d$. In this case all excitons describe a single $p$-excitation  delocalized over the chain, with spatial distribution reminiscent of the eigenstates of a quantum particle in a 1D box.
Dislocations or disorder, however, lead to more localized exciton states \cite{cenap:motion,eisfeld:disorder}, shown in \fref{chain_excitons} (bottom).

Due to the $|r_n-r_m|^{-3}$ dependence of dipole-dipole interactions, nearest neighbor interactions $J=C_3/d^3$ are dominant in \bref{Hdd}. Neglecting longer range interactions,  we can analytically solve the eigenvalue problem \bref{eigensystem} to obtain 
\begin{eqnarray}
U_k = -2 J \cos{[\pi k /(N+1)]},\\
\ket{\varphi_k} =\sqrt{2/(N+1)}\sum_{n=1}^N  \sin{[\pi k n /(N+1)]} \ket{\pi_n},
\label{analytical_excitons}
\end{eqnarray}
with $k=1,\cdots ,N$.

For large $N$, the energies become densely spaced and form the exciton band, with bandwidth 2J. Eigenstates as shown in \fref{chain_excitons} can be experimentally accessed using appropriately detuned microwave pulses, see \cite{moebius:cradle}. However this works straightforwardly only for the most symmetric states ($n=1$ in the figure), since the microwave will act symmetrically on all atoms and thus provides no transition matrix element towards asymmetric states such as $n=5$.  

\begin{figure}[htb]
\centering
\includegraphics[width= 0.8\columnwidth]{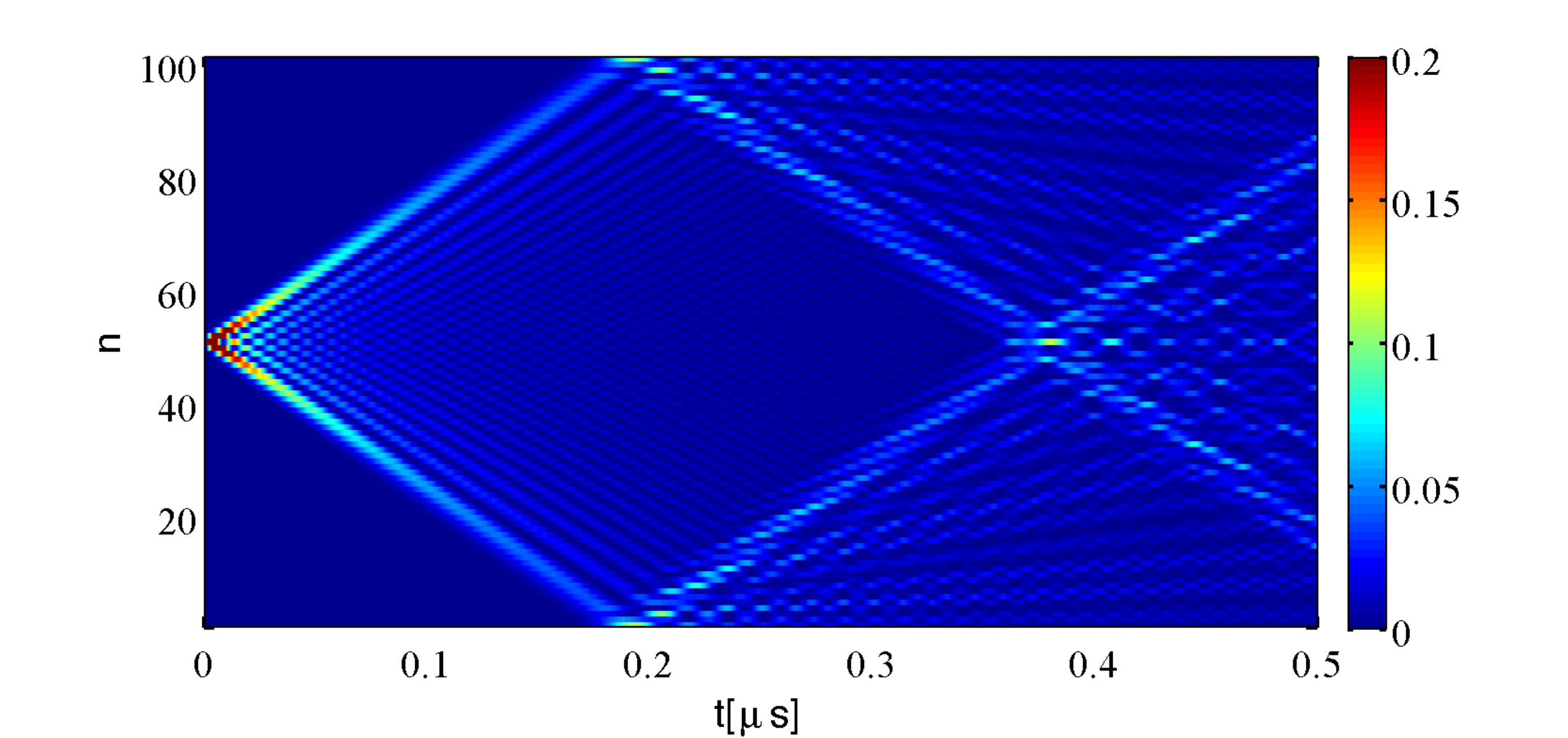} 
\caption{Ballistic transport on a large static Rydberg aggregate with $N=100$ Rubidium atoms at $\nu=80$, with atoms numbered by $n$. These are separated by $d=8\mu$m (along the vertical axis). The excitation probability per atom $p_n=|\braket{\pi_n}{\Psi(t)}|^2$ is shown as color shade.  Transport is fast and occurs within the overall effective life-time of $2\mu$s and prior to any acceleration.
\label{ballistic_transport}}
\end{figure}

A static Rydberg aggregate initialized with an excitation localized on just one specific atom is not in one of the eigenstates \bref{analytical_excitons}. The corresponding time-evolution in a superposition of eigenstates then leads to ballistic quantum transport as shown in \fref{ballistic_transport}, using \eref{Tully_electron} with fixed atomic positions.

A more natural scenario for Rydberg atom placement in an ultracold gas environment are random positions, with the exclusion of close proximities due to the dipole blockade \cite{jaksch:dipoleblockade,lukin:quantuminfo}. In this case, some excitons naturally localize on the clusters with the most closely spaced atoms, similarly to \fref{chain_excitons} (bottom).

For more complex assemblies of sites (atoms) than the simple linear chain in \fref{ballistic_transport}, the coherent quantum dynamical propagation there can also be viewed as a quantum random walk \cite{muelken:slowtransp,muelken:CTRW_dendrimer,muelken:quantumwalks:review}.

 \subsection{Further effects of disorder}\label{disorder}

Even a static Rydberg aggregate will not show the perfectly regular structure assumed in the simulation for \fref{ballistic_transport}. The position of the atoms will be subject to a certain degree of randomness, either because atoms are  randomly placed within a gas, or because of quantum or thermal  fluctuations even in case of well defined individual trap centers. Since fluctuations in the position affect the interaction strengths in \bref{Hdd}, this leads predominantly to \emph{off-diagonal disorder}.

Another possible disorder is \emph{diagonal disorder}, if excitations located on different sites have slightly different on-site energies $E_n$, captured in the addition to the Hamiltonian
\begin{equation}
\sub{\hat{H}}{dis}(\bv{R})=\sum_{n}^N E_n  \ket{\pi_n}\bra{\pi_n}.
\label{Hdis}
\end{equation}
In Rydberg aggregates this is typically negligible, but can still be present due to inhomogeneous stray fields which may lead to different differential Stark or Zeeman shifts between $\ket{s}$ and $\ket{p}$ for different members of the aggregate. 

It is known that in large systems, disorder typically leads to the localization of excitations \cite{fidder:molagg:disorders}. For the special case of, typically rather smaller, Rydberg aggregates, the consequences of filling disorder have been investigate in small ordered 2D grids \cite{yu:smallgrid:excithopping} and off-diagonal disorder in 2D and 3D lattices in \cite{Robicheaux:randompos:dipdip}.

Diagonal disorder in molecular aggregates often follows a general L{\'e}vy-distribution. The resulting frequent occurrence of large outliers in the energy of certain sites, can give rise to additional localization features \cite{bas:levy,vlaming:disorder,eisfeld:disorder,moebius:exciton_levy:arxiv}.

\subsection{Experimental realizations}\label{static_realisations}

The static Rydberg aggregates described so far have recently been experimentally realized in two configurations.

\begin{figure}[htb]
\centering
\includegraphics[width= 0.8\columnwidth]{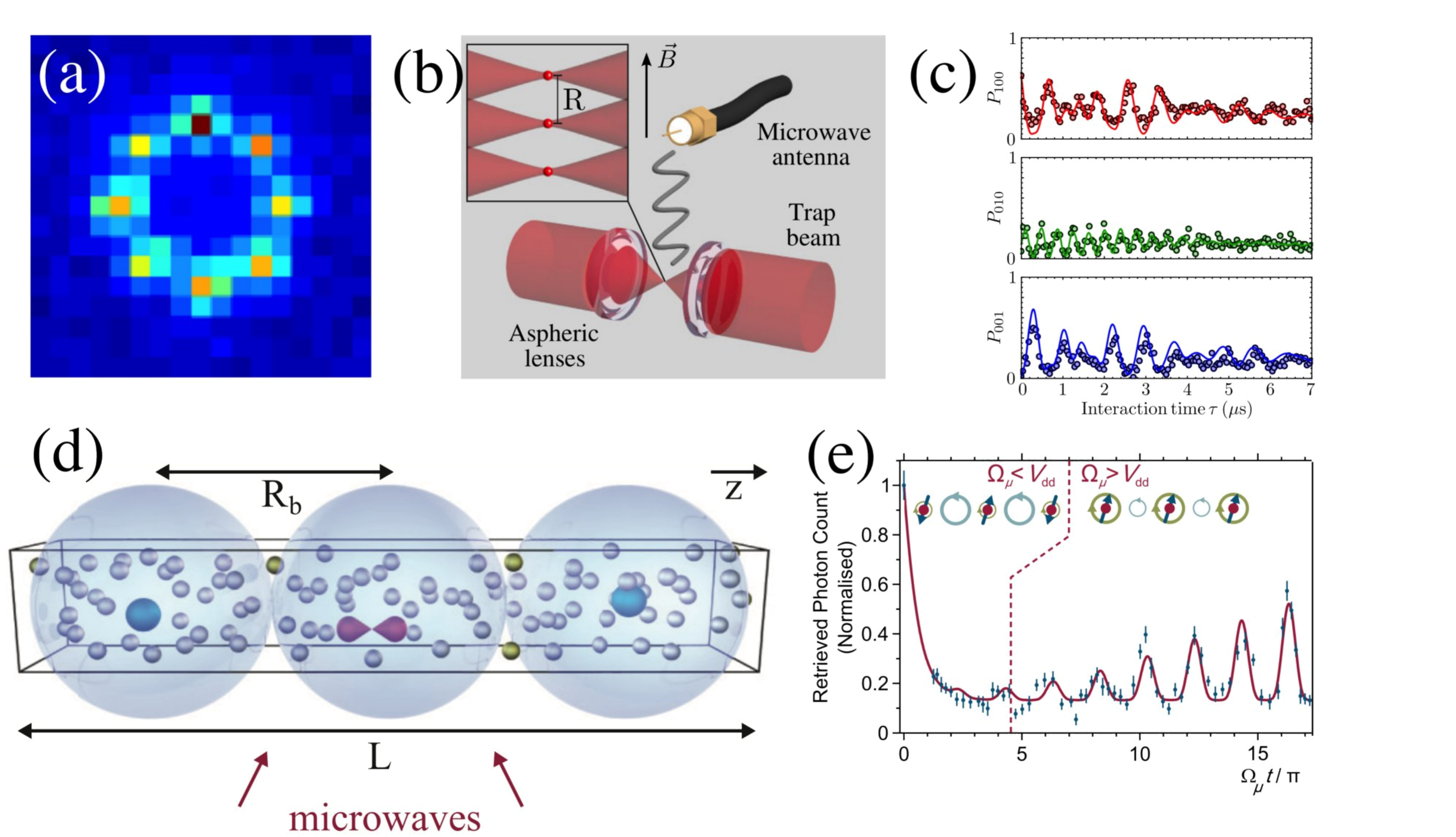} 
\caption{Recent experimental progress towards Rydberg aggregates. (a) Fluorescence image of trapped ground-state atoms in a ring arrangement, diverse 2D geometries are feasible [from \cite{nogrette:hologarrays}]. This is achieved with optical dipole trap arrays as shown in (b) for a trimer aggregate. (c) Quantum transport, analogous to \fref{ballistic_transport}, on the trimer aggregate in (b) [from \cite{barredo:trimeragg}]. Alternative creation of an emerging Rydberg aggregate (blue/violet balls) within a bulk cold gas (yellow/grey balls) [from \cite{cenap:emergent}]. (e) Interplay of excitation transport and microwave drive, see text [from \cite{maxwell:polaritonstorage}].
\label{experimental_aggregates}}
\end{figure}

{\it Regular array: }To create a regular Rydberg array, cold ground-state atoms were first loaded into a lattice of optical dipole-traps, which can be arranged in almost arbitrary 2D geometries through the use of spatial light modulators \cite{nogrette:hologarrays}. These atoms are then promoted to $\ket{62d}$ and $\ket{63p}$ Rydberg states.
 Exciton dynamics on a dimer \cite{ravets:foersterdipdip} and a small regular trimer of Rydberg atoms could be recorded precisely \cite{barredo:trimeragg}, as reprinted in \frefp{experimental_aggregates}{c}, with different panels showing the excitation population on aggregate atom $k=1,2$ and $3$. The graphs correspond to cuts of \fref{ballistic_transport} for $N=3$ at the respective site of the atoms. 
 The traps are disabled during excitation and exciton transport, but short time-scales ensure that Rydberg atom motion is only a small disturbance, see also \sref{interactions}. These techniques have recently been extended to larger arrays studying spin systems \cite{labuhn:rydberg:ising}.

{\it Bulk excitation: }If Rydberg excitation takes place directly in a bulk cold gas, without prior single atom trapping, the positions of the aggregate atoms will be a-priori unknown and disordered. This was the starting point in  \cite{cenap:emergent,maxwell:polaritonstorage}, albeit for experiments with a focus on photon-atom interfacing. These also contained small numbers of Rydberg excitations, $N\approx 3$.  
In these experiments excitation transport such as displayed in \fref{ballistic_transport} competes with microwave driving between the two coupled atomic states, as shown in \frefp{experimental_aggregates}{e}.  When dipole interactions are stronger than the driving ($V_{dd}>\Omega_\mu$ in the figure), they can decohere a photon wavepacket stored in an EIT medium, thus reducing the retrieved photon count plotted. Experiments take place on sub microsecond time-scales, thus atomic motion is negligible. The article \cite{cenap:emergent} contains a theoretical analysis of these experiments in terms of emergent Rydberg lattices (aggregates).

Another experiment with disordered bulk Rydberg aggregates is covered in \rref{guenter:EITexpt}. Here the second Rydberg angular momentum state is introduced through a F{\"o}rster resonance of the type $\ket{ss'}\rightarrow\ket{pp}$. In the resulting assembly of $N\sim 10-100$ Rydberg atoms, containing both $s$ and $p$ states, excitation transport again takes place through resonant dipole-dipole interactions. In that experiment, the resulting spatial diffusion of $p$ states can be monitored and controlled through an imaging scheme exploiting the optical response of the cold gas in the background \cite{guenter:EIT}. Also in this system, competition between microwave drive and dipole exchange can lead to interesting relaxation dynamics \cite{pineiro_orioli_spinrelax}.

{\it Excitation facilitation: }Complementary to coherent Rydberg aggregates on which we focus here, there are aggregates formed incoherently through a competition of Rydberg excited state decay and continuous excitation. On time-scales large compared to the Rydberg life-time, the decoherence introduced by spontaneous excited state decay gives rise to dynamics that 
can often be described by classical rate equations \cite{cenap:manybody}. Using these, one can predict the ``aggregation'' (facilitated excitation) of assemblies of Rydberg atoms with nearest neighbor spacing $d$   \cite{lesanovsky:nonequil_structures,gaertner:faciitationagg:theory}, for excitation laser detuning $\Delta$ that causes resonance for double excitation at a preferred distance $d=[C_6/(2\Delta)]^{1/6}$, where $C_6$ is the van-der-Waals coefficient. This further manifestation of excitation anti-blockade \cite{cenap:antiblockade}, was subsequently observed in a variety of experiments \cite{malossi:fullcountstat,Simonelli:excitationavalanche:JPB,valado:exp:kineticconstraints,schempp:countingstat,urvoy:vaporaggreg}. Such systems show 
parallels to the physics of glassy systems \cite{lesanovsky:kinetic}. Even these incoherently created Rydberg aggregates can form the basis for coherent excitation transport as reviewed here, if augmented at some moment with the coherent introduction of a single $p$-excitation and then monitored for short times.

\subsection{Motional decoherence}\label{decoherence}

The Rydberg aggregate experiments cited in \sref{static_realisations} do not constrain positions of the atoms, but take place on time scales before significant motion 
occurs. Nonetheless residual thermal atomic motion has some effect, as discussed in \cite{barredo:trimeragg} and its supplemental material. In terms of the previous section this motion can be viewed as  leading to time-dependent off-diagonal disorder.

Hence, motion may be regarded as a noise source, or if controllable e.g.~through temperature, as a tuneable asset. In the next section we consider an alternative twist, with which atomic motion can be turned from a hindrance into an interesting feature.

\section{Flexible Rydberg aggregates}\label{flexible}

Static Rydberg aggregates reviewed above are an interesting model platform for the study of excitation transport, spin models and the impact of disorder. However, 
residual atomic motion represents a source of decoherence or noise. Turning this negative viewpoint around, we now summarize the rich physics arising from 
fully embracing directed or controllable motion of all atoms constituting a \emph{flexible} Rydberg aggregate. We imply with the term ``flexible'' that atoms do not have a preferred (equilibrium) position along directions in  which they are free to move.
The motion can, and for conceptual simplicity should, still be confined to a restricted geometry, such as provided by a one-dimensional optical trap.
\begin{figure}[htb]
\centering
\includegraphics[width= \figurewidth\columnwidth]{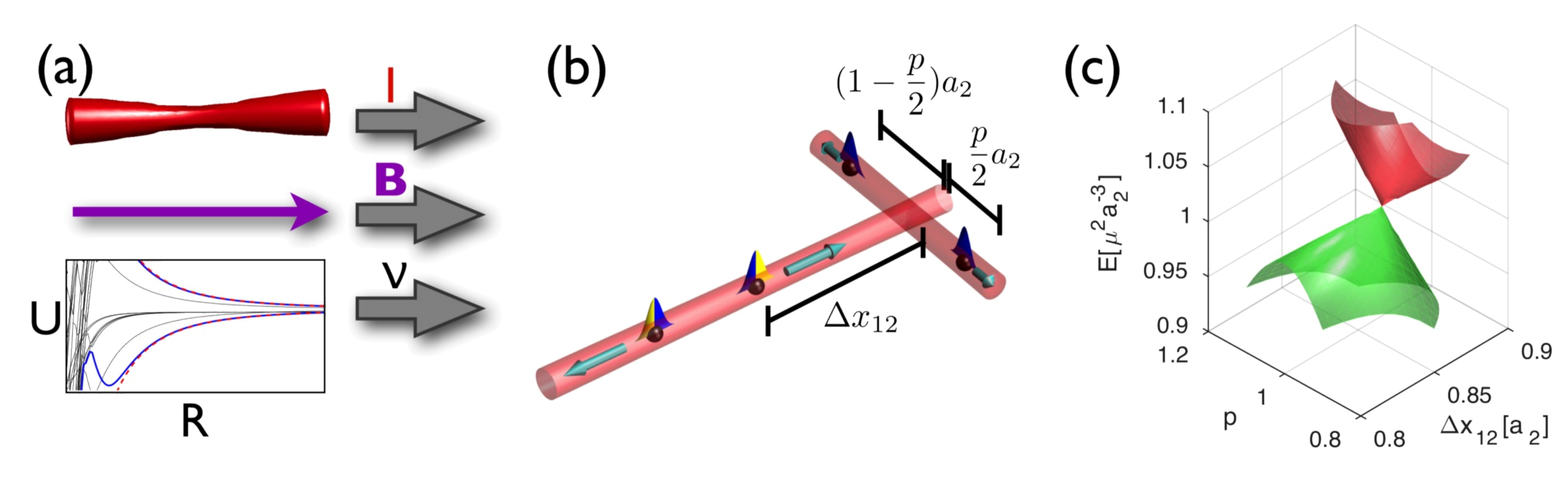} 
\caption{Tailoring Rydberg aggregate Born-Oppenheimer surfaces. (a) Atomic dynamics depends strongly on principal Rydberg quantum number $\nu$, chosen distance scale $R$, external field (e.g.~magnetic fields $\mathbf{B}$) for tuning interactions and isolating electronic states as well as the design of the atomic trapping (e.g.~intensity $I$ of a red detuned laser). (b) Together these components constrain and tune atomic motion possible in a flexible Rydberg aggregate. (c) Born-Oppenheimer surfaces can thusly be designed to contain, for example, accessible conical intersections [adapted from \cite{leonhardt:switch}], for geometrical variables $p$, ${\Delta}x_{12}$, $a_2$ see (b).
\label{flexible_tailoring}}
\end{figure}
%

\subsection{Tailored Born-Oppenheimer surfaces}\label{taylored_surfaces}

We have discussed in \sref{excitons_surfaces}, how a compact essential states model for each of the Rydberg atoms allows simple calculations of the excitonic Born-Oppenheimer surfaces that govern the motion of the aggregates. In contrast to the situation within normal, tightly-bound molecules, the long range Rydberg-Rydberg interactions create potential landscapes with characteristic length scales of micrometers. These scales exceed the minimal ones on which additional external potentials from magnetic or optical trapping can act on the atoms. All together, the ingredients sketched in \fref{flexible_tailoring} allow the tailoring of Born-Oppenheimer surfaces governing a Rydberg aggregate to provide atomic motional dynamics with interesting and diverse properties. In the following we review the properties already studied and provide an outlook on future opportunities.

\subsection{Adiabatic excitation transport}\label{flexible_transport}
\label{cradle}

\begin{figure}[htb]
\centering
\includegraphics[width=0.8\columnwidth]{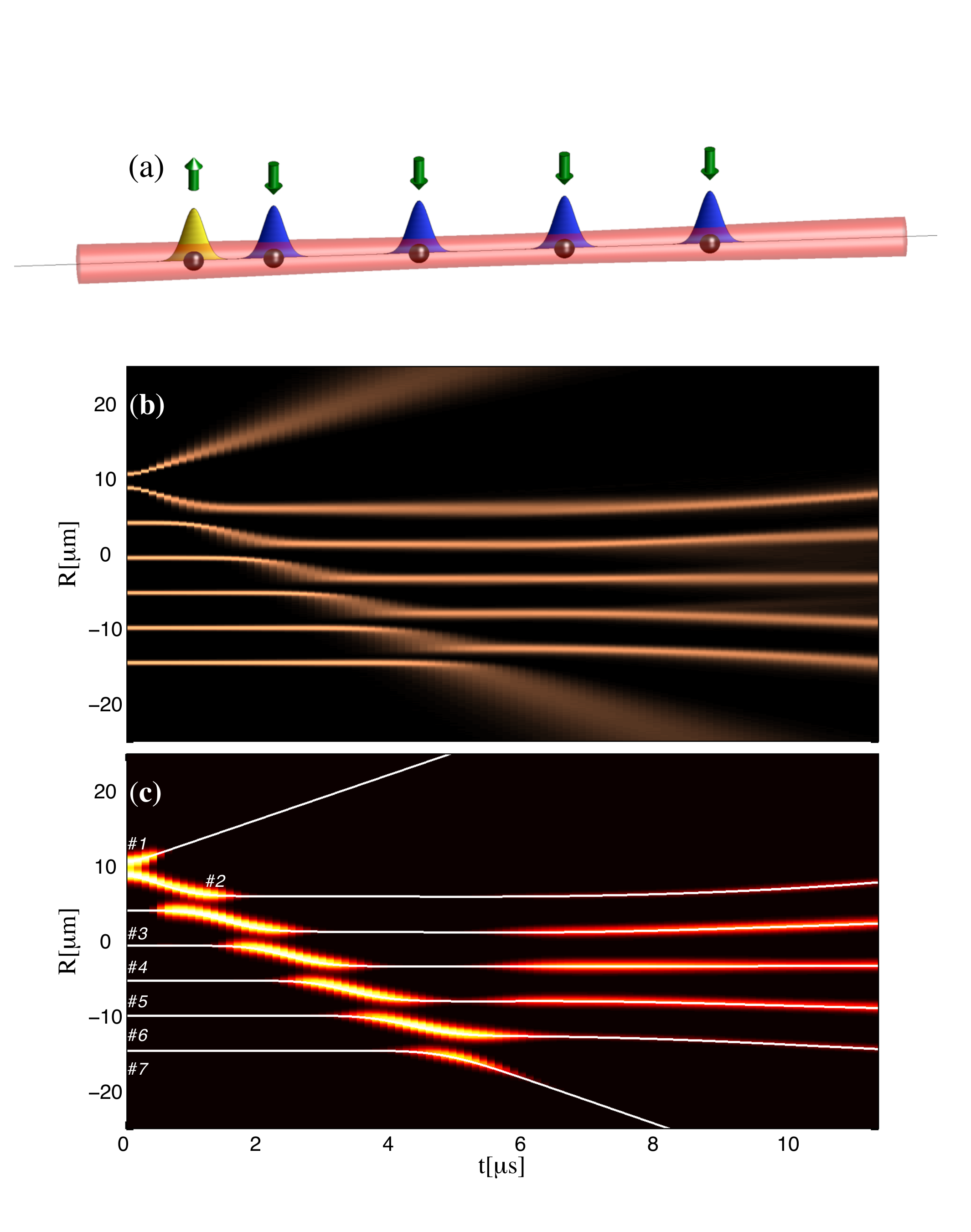} 
\caption{Adiabatic excitation transport in a one-dimensional flexible Rydberg aggregate, as discussed in \cite{wuester:cradle,moebius:cradle}. (a) Visualization of setup, atoms in $\ket{s}$ (blue) and $\ket{p}$ (yellow) are shown with their position uncertainty (shaded line) and optical trap (red). We also sketch the corresponding pseudo-spin representation. (b) Scaled total atomic density showing successive elastic collisions as in Newton's cradle. (c) Corresponding excitation probability for atoms $1$-$7$. White lines show the mean position of each atom, and the color shading and its width indicate the excitation probability $p_n(t) = |\braket{\pi_n}{\Psi(t)}|^2$. Panels (b,c) are adapted from \cite{wuester:cradle}.
\label{newtons_cradle}}
\end{figure}
Already the simplest flexible aggregate, with $N$ atoms free to move in a one-dimensional optical trap, is a promising candidate to study the interplay of excitation transport and motional dynamics expected from the discussion in \sref{models}. Consider the scenario of an equidistant chain of atoms in a one-dimensional optical trap, with a dislocation at the end, visualized in \frefp{newtons_cradle}{a}. The dislocation causes initial close proximity of two atoms. In this situation, one of the BO surfaces of \eref{eigensystem} $\sub{U}{rep}(\bv{R})$ corresponds to the single p-excitation  localized on the two dislocated atoms initially. This feature is also evident in the randomly dislocated chain analyzed in \fref{chain_excitons} (bottom).
The surface $\sub{U}{rep}(\bv{R})$ then causes strong repulsion of  the dislocated atoms only, leading to motional dynamics with successive repulsive pairwise collisions between all atoms in the chain \cite{wuester:cradle,moebius:cradle}, shown in \frefp{newtons_cradle}{b}. Interestingly, this classical momentum transfer between the atoms is accompanied 
by a  directed and coherent quantum transport of the single electronic excitation through the chain with high fidelity as shown in \frefp{newtons_cradle}{c}. Further analysis reveals that the underlying exciton state $\ket{\sub{\varphi}{rep}(\bv{R})}$ is adiabatically preserved due to the relatively slow motion $\bv{R}(t)$ of the atoms. In this scenario, the coherent excitation transport also implies entanglement transport, which could be measured through Bell-type inequalities \cite{wuester:cannon}. 

An additional combination of repulsive van-der-Waals and attractive dipole-dipole interactions can modify exciton states at short distances and form aggregates of actually up to 6 atoms \emph{bound} in one dimension, enabling studies of their dissociation \cite{zoubi:VdWagg}. With additional Stark tuning of Rydberg levels, such a binding mechanism can even hold in 3D for dimers and trimers\cite{kiffner:dipdipdimer,kiffner:fewbodybound}.

\subsection{Conical intersections}\label{conical_intersections}
\begin{figure}[htb]
\centering
\includegraphics[width=0.8\columnwidth]{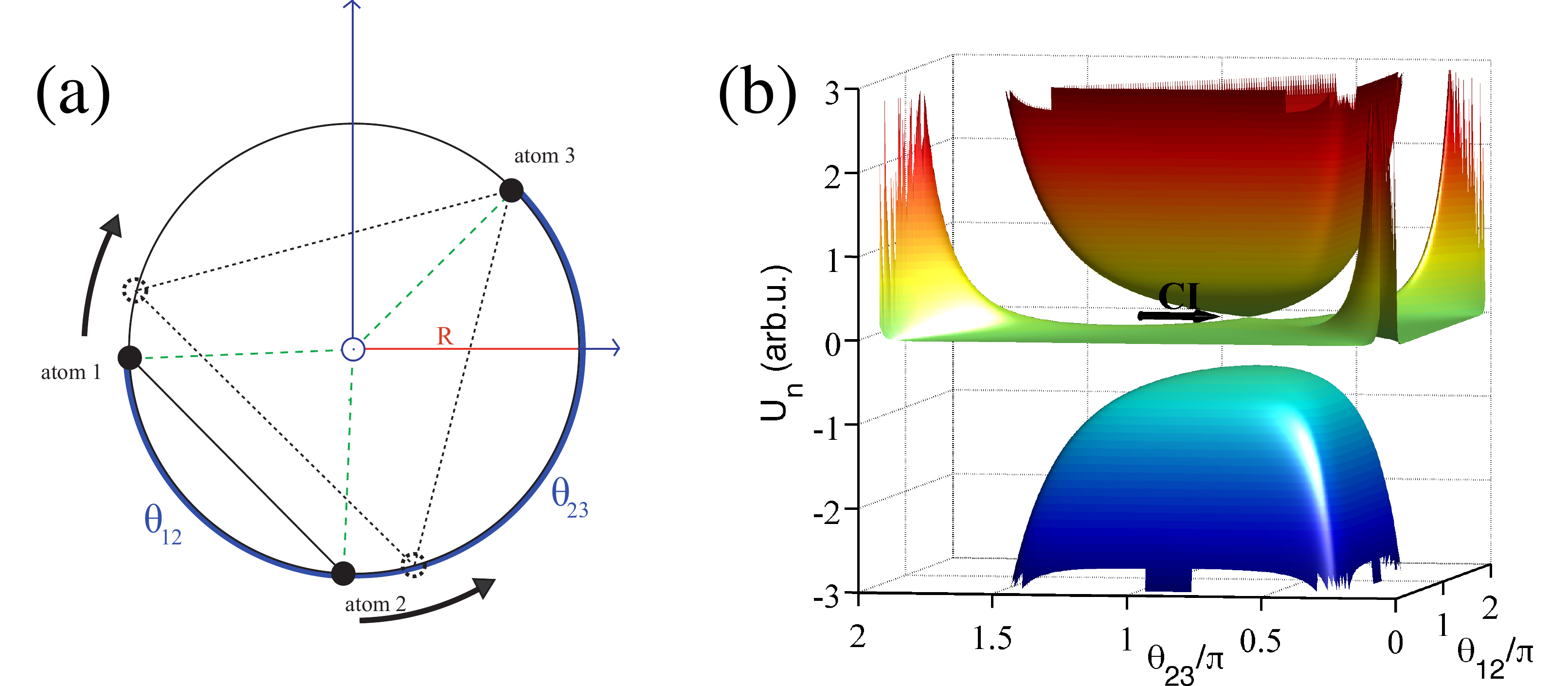} 
\caption{(a) Geometry of Rydberg Ring trimer. Atomic positions are parametrised in terms of the angles $\theta_{12}$, $\theta_{23}$. (b) The three BO surfaces following from \eref{eigensystem}, as a function of the two relative angles indicated in (a). The conical intersection is marked between the upper two. Figure adapted from \cite{wuester:CI}.
\label{conical_intersection}}
\end{figure}

%
In  the preceding section we have seen that Rydberg atomic motion on a single BO surface can already lead to intriguing transport physics far from being trivial. Additional and qualitatively different features come into play if multiple surfaces are involved, for example due to the presence of conical intersections (CIs) \cite{yarkony1996diabolical,yarkony2001conical}. They generically occur in dipole-dipole interacting aggregates described by \bref{Hdd} whenever the motion - which still may be  geometrically restricted - takes place in two or more spatial dimensions \cite{wuester:CI}. The simplest example is a ring timer. It possesses one well localized conical intersection (seam) of surfaces, as drawn in \fref{conical_intersection}. As a consequence, we expect  a Rydberg atom wavepacket, which traverses the CI, to undergo  non-adiabatic transitions. This also should lead to optically resolvable manifestations of Berry's phase \cite{wuester:CI}.

We can study the impact of CIs on excitation transport, as discussed in \sref{cradle}, in a scenario with two intersecting linear chains. Then, the CI can be functionalized as a switch, where minute changes in aggregate geometry have profound impact on the exciton transport direction and coherence properties \cite{leonhardt:switch}. 

\begin{figure}[htb]
\centering
\includegraphics[width=0.99\columnwidth]{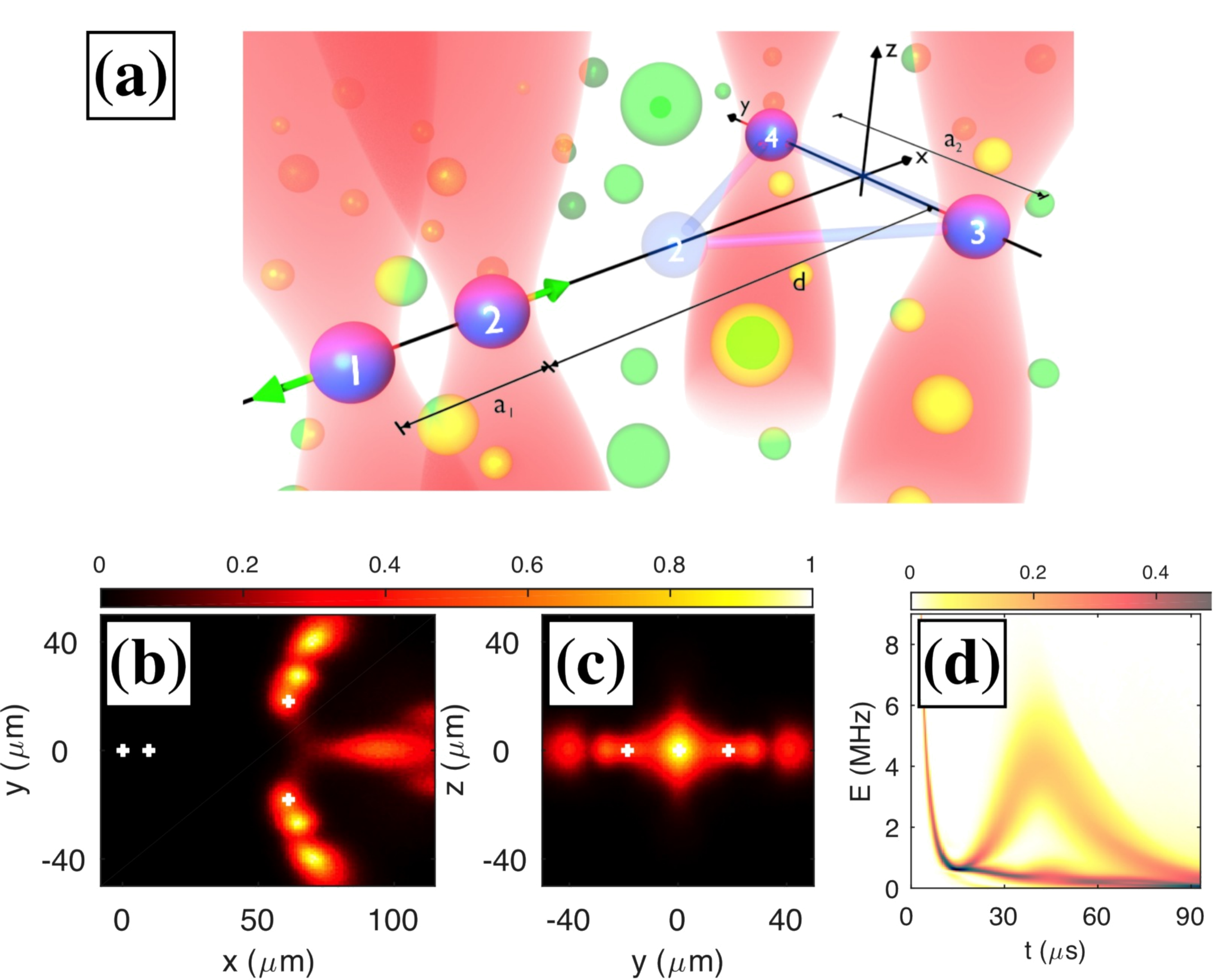} 
\caption{Manifestation of non-adiabatic motional dynamics for atoms moving freely in three dimensions. (a) Sketch of initial geometry, with atoms one and two in close proximity on the $x$-axis, and three and four somewhat wider spaced on the $y$ axis. Atoms one and two will initially repel strongly along the green arrows.
(b,c) Ensuing two-dimensional column densities of atoms at late times, showing several lobes due to multiple populated BO surfaces. White crosses mark the initial positions of atoms $1-4$ as shown in panel (a). (d) Time-evolution of potential energy distribution. The splitting on two BO surfaces can also clearly be seen here. Figures taken from \cite{leonhardt:unconstrained}.
\label{embedded_intersection}}
\end{figure}

Finally non-adiabatic features can also be accessed entirely without atomic confinement, by choosing suitable initial Rydberg excitation locations in a 3D cold gas volume, with Rydberg atoms subsequently moving without constraints \cite{leonhardt:unconstrained}. These results, presented in \fref{embedded_intersection}, move an experimental demonstration of non-adiabatic effects due to passage through a conical intersection in Rydberg aggregates firmly within reach of present day technology \cite{balewski:elecBEC,guenter:EITexpt,thaicharoen:trajectory_imaging,celistrino_teixeira:microwavespec_motion}. 

For free motion in 3D, a more sophisticated model for dipole-dipole interactions than \bref{Hdd} is needed in general, which takes into account azimuthal quantum numbers $m_l$ of the electronic states and the anisotropy of dipole-dipole interactions. 
This was used in \cite{leonhardt:unconstrained}. There are no qualitative changes in the type of features observed, compared to the simple isotropic model described above. In fact, by isolating specific $m_l$-states with Zeeman shifts, one can tune experiments such that the isotropic model directly applies \cite{leonhardt:orthogonal}.

Other interesting viewpoints on conical intersections within dipole-dipole energy surfaces of Rydberg dimers are highlighted in \cite{kiffner:spinorbitcoupl,kiffner:nonabelian}, for dimers bound as in \cite{kiffner:dipdipdimer,kiffner:fewbodybound}: The non-adiabatic coupling terms in \bref{nonadiab} can be assembled into a vector potential, see e.g.~\cite{Zygelman:nonabelian,domke:yarkony:koeppel:CIs}, to which the motion in \bref{fullSE_adiab} is then coupled. The system thus furnishes a platform for synthetic spin-orbit coupling \cite{kiffner:spinorbitcoupl} and non-Abelian gauge fields \cite{kiffner:nonabelian}.

%

\subsection{Experimental realizations of unfrozen Rydberg gases}\label{flexible_realisations}

Following our summary of theory proposals for the applications of flexible Rydberg aggregates, we now present a brief overview of the current capabilities of experiments in that direction.
Crucial to the implementation is a sufficiently controlled positioning and equally importantly, high resolution and time resolved read-out of atom locations.

Many recent Rydberg atomic experiments are geared towards utilizing the dipole-blockade for quantum computation, quantum optics or simulation of condensed matter systems. 
For these objectives motion represents a noise source (see \sref{decoherence}) and hence is suppressed by design. 

\begin{figure}[htb]
\centering
\includegraphics[width= 0.9\columnwidth]{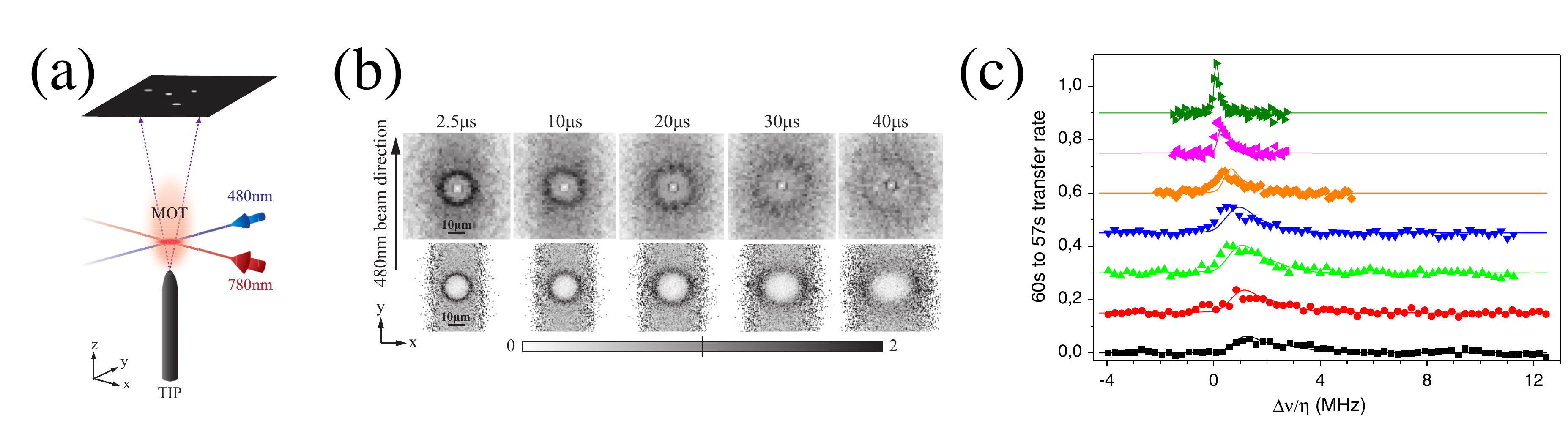} 
\caption{Experimental techniques for the observation of flexible Rydberg aggregates. (a) Diagram for spatially resolved field ionization. (b) Rydberg pair correlations observed in this manner as a function of time. Experiment (top) and simulations (bottom) both show the movement of the correlation peak to larger separations due to dipole dipole repulsion. (a,b) taken from 
\cite{thaicharoen:trajectory_imaging}. (c) Microwave spectroscopy of Rydberg atoms in motion. The interaction induced broadening of a resonance signal that is present initially (bottom) reduces as atoms repel and interactions diminish (towards top). Reproduced from \cite{celistrino_teixeira:microwavespec_motion}.
\label{flexible_experimental}}
\end{figure}
Early experiments had a stronger fundamental atomic physics focus, studying atomic interactions, the resulting collisions and ionization \cite{Fioretti:longrangeforces,park:dipdipionization,li_gallagher:dipdipexcit} or the autoionization of Rydberg atoms when these are set into motion due attractive van-der-Waals interactions \cite{amthor:autoioniz,Amthor:mechVdW,amthor:modellingmech,Faoro:VdWexplosion}.
Recent experiments demonstrated the continuous observation of Rydberg atoms in cold gases while these undergo motional dynamics due to van-der-Waals interactions. The first technique used to this end, is time- and space resolved atomic field ionization \cite{thaicharoen:trajectory_imaging,thaicharoen:dipolar_imaging}. Spatial resolution is provided by strong electric field in-homogeneities near a tip, which enable the reconstruction of the point of origin of ion-counts on a detector, as sketched in \frefp{flexible_experimental}{a}. The correlation signals observed, see \frefp{flexible_experimental}{b}, show the effect of vdW repulsion, and could similarly give insight into motional aggregate dynamics such as that shown in \fref{newtons_cradle} or \fref{embedded_intersection}.

The second technique relies on microwave spectroscopy \cite{celistrino_teixeira:microwavespec_motion},  monitoring the time-dependence of two-photon microwave transfer rates from one Rydberg state to another. The signal is significantly broadened due to vdW interactions of the Rydberg atoms initially, but narrows considerably as the atom cluster undergoes vdW repulsion and potential energies are reduced, see \frefp{flexible_experimental}{c}. Similar techniques can measure the time-dependent BO surface energy \eref{eigensystem} of a flexible Rydberg aggregate,  as shown in \frefp{embedded_intersection}{d}.

\section{Embedded Rydberg aggregates}\label{embedded}

The most common starting point for experiments is to excite few cold atoms within a host cloud to Rydberg states, typically with angular momenta $s$ or $d$ such that they can be reached via a two-photon excitation. Then, the dipole-blockade guarantees a Rydberg assembly with a minimal nearest neighbor distance equal to the blockade radius, roughly $r_{bl}=(C_6/\Omega)^{1/6}$, where $\Omega$ is the Rabi-frequency of excitation. Augmenting  the assembly with one or few atoms in an adjacent angular momentum state (e.g.~$p$) creates a Rydberg aggregate. 
The augmentation can be achieved via microwave coupling \cite{maxwell:polaritonstorage,pineiro_orioli_spinrelax}, or direct laser excitation \cite{weber:superatoms}.

One might expect that the the embedding cold background  gas will strongly modify the
dynamics of flexible or static Rydberg aggregates created in such a manner. 
Interestingly, there should be parameter regimes  where this is not the case  \cite{leonhardt:unconstrained,moebius:bobbels}, which is experimentally corroborated in \cite{celistrino_teixeira:microwavespec_motion,thaicharoen:trajectory_imaging}, where observations on Rydberg atoms moving through a gas could be explained with classical trajectory simulations containing only the Rydberg excited atoms, without direct or indirect influence of the hosting cold gas cloud. 

However, for other parameter regimes strong life-time reductions of high Rydberg excitations \cite{balewski:elecBEC,schlagmueller:ucoldchemreact:prx} or their associative ionization have been reported \cite{niederpruem:giantion}. Additional processes that arise in the system are the formation of ultra-long range molecules, see e.g.~\cite{greene:ultralongrangemol,Bendkowsky:boundquantumreflection,Bendkowsky:ultralongmol,Li:homonuc}, polarons \cite{Schmidt:rydpolaron}, and Rydberg molecule induced spin flips \cite{niederpruem:remotespinflip}. If by experimental design these effects can be made sufficiently small, aggregate physics in a cold gas environment proceeds roughly as discussed so far. Individual atoms in a Rydberg assembly can even be replaced by super-atoms (groups of ground-state atoms coherently sharing a Rydberg excitation), without changing the results described \cite{moebius:bobbels}.

If a Rydberg aggregate can be embedded in an ultracold gas environment without detriment, the latter can actually be used to some advantage. It can serve as an atomic reservoir for repeated excitation of Rydberg atoms \cite{wuester:cannon} but also as a gain medium for sensitive position- and state detection of Rydberg excitations  \cite{olmos:amplification,guenter:EIT,guenter:EITexpt,schoenleber:immag}. These schemes add to the toolkit that promises sufficient temporal and spatial resolution for the experimental readout of the processes summarized earlier. Even more interestingly, they provide a control knob over decoherence, which is created by the measurement or readout within the system. It can give rise to decoherence between electronic states forming a Born-Oppenheimer surface, causing a complete cessation of intra-molecular forces \cite{wuester:immcrad}. It also decoheres excitation hopping such as shown in \fref{ballistic_transport}, enabling quantum simulations of light harvesting \cite{schoenleber:immag}, switchable Non-Markovianity \cite{genkin:markovswitch}, tuneable open quantum spin systems \cite{schempp:spintransport} and engineered thermal reservoirs \cite{schoenleber:thermal}.

\section{Rydberg dressed aggregates}\label{dressed}

\begin{figure}[htb]
\centering
\includegraphics[width=0.6\columnwidth]{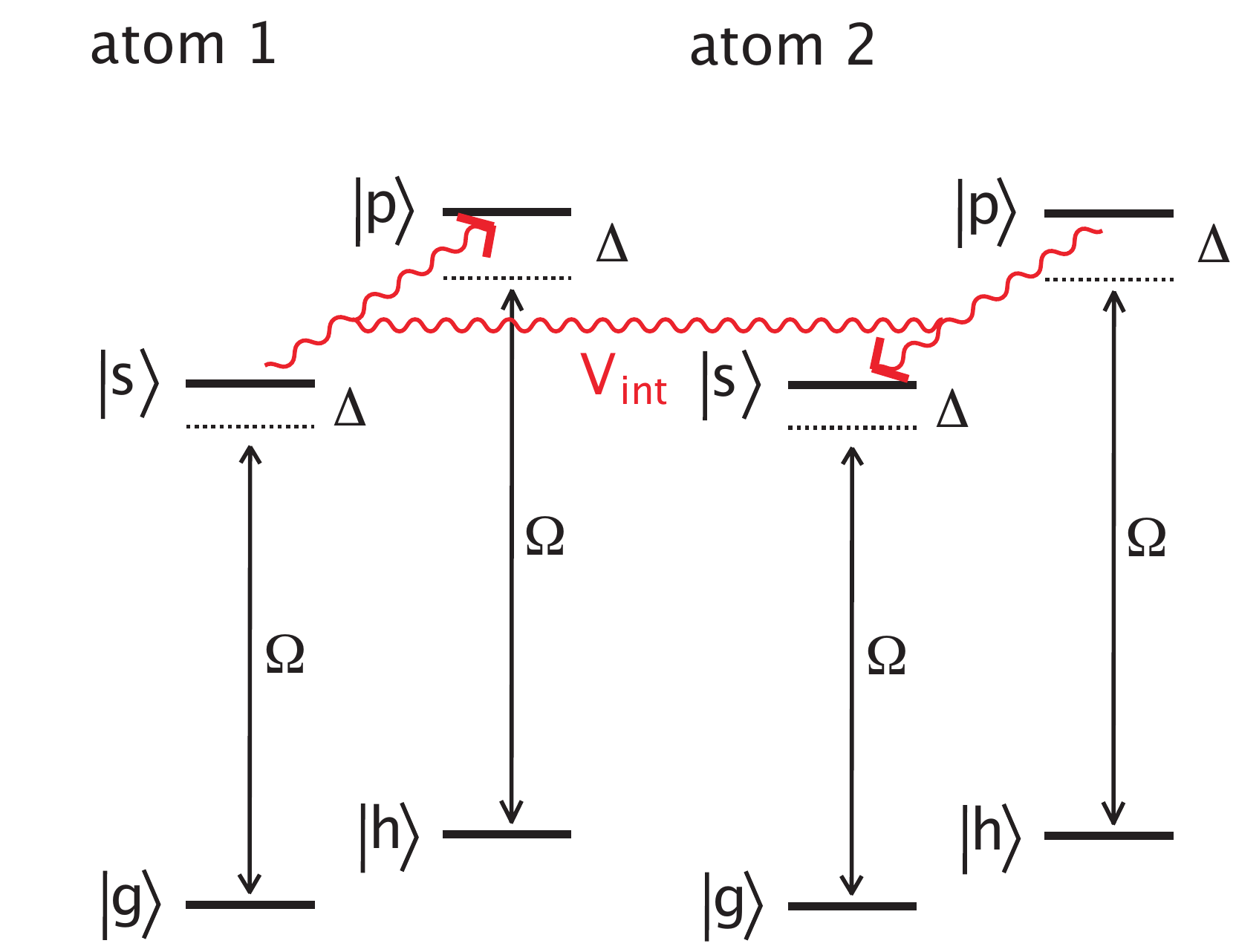} 
\caption{Level scheme of dressed dipole-dipole interactions. Separate couplings with identical effective Rabi frequency $\Omega$ enable transitions $\ket{g}\leftrightarrow\ket{s}$ and $\ket{h}\leftrightarrow\ket{p}$, respectively. Both are far detuned $\Omega \ll \Delta$. Thus, while atoms predominantly remain in the ground states $\ket{g}$, $\ket{h}$, the small admixture of Rydberg levels allows an effective state-change interaction $\hat{H}\sim \ket{gh}\bra{hg}$.
\label{dressed_leveldiag}}
\end{figure}

Most phenomena reviewed so far can be ported to different parameter regimes, if the interactions discussed in \sref{basicmodel} are replaced by \emph{dressed dipole-dipole interactions} \cite{wuester:dressing}. A direct mapping of resonant dipole-dipole interactions requires two separate ground states $\ket{g}$ and $\ket{h}$, that are far off resonantly coupled to the two Rydberg states $\ket{s}$ and $\ket{p}$ employed here earlier, as sketched in \fref{dressed_leveldiag}. This can effectively give rise to a state-flip interaction Hamiltonian $\hat{H}\sim (\ket{gh}\bra{hg} + \mbox{c.c})$ in the space of the ground-states \cite{wuester:dressing}.

While making an experiment more challenging, the use of dressed interactions gives additional flexibility for the choice of parameters. For example the ring-trimer CI shown in \fref{conical_intersection} can be realized with dressed interactions \cite{wuester:CI} to avoid the need for extreme trap strengths. Dressing can also match the energy/time-scales of Rydberg physics with those of ground-state cold atom- or Bose-Einstein condensate physics.

Moreover, dressed interactions can  enable entirely novel effects not present in systems with bare interactions. They can give rise to the dipole-dipole induced repulsion of droplets containing many atoms \cite{genkin:dressedbobbles}, even allowing them to be brought into quantum superpositions of different droplet locations \cite{moebius:cat}. The droplets  are then described by mesoscopically entangled states with tens of atoms. Dressed dipole-dipole interactions could also be exploited to design quantum simulators of electron-phonon interactions \cite{hague:su:schrieffer:heeger,hague:rydquantsim,hague:supercondquantsim}.

\section{Rydberg aggregate existence domains}\label{interactions}

%
\begin{figure}[htb]
\centering
\includegraphics[width= \figurewidth\columnwidth]{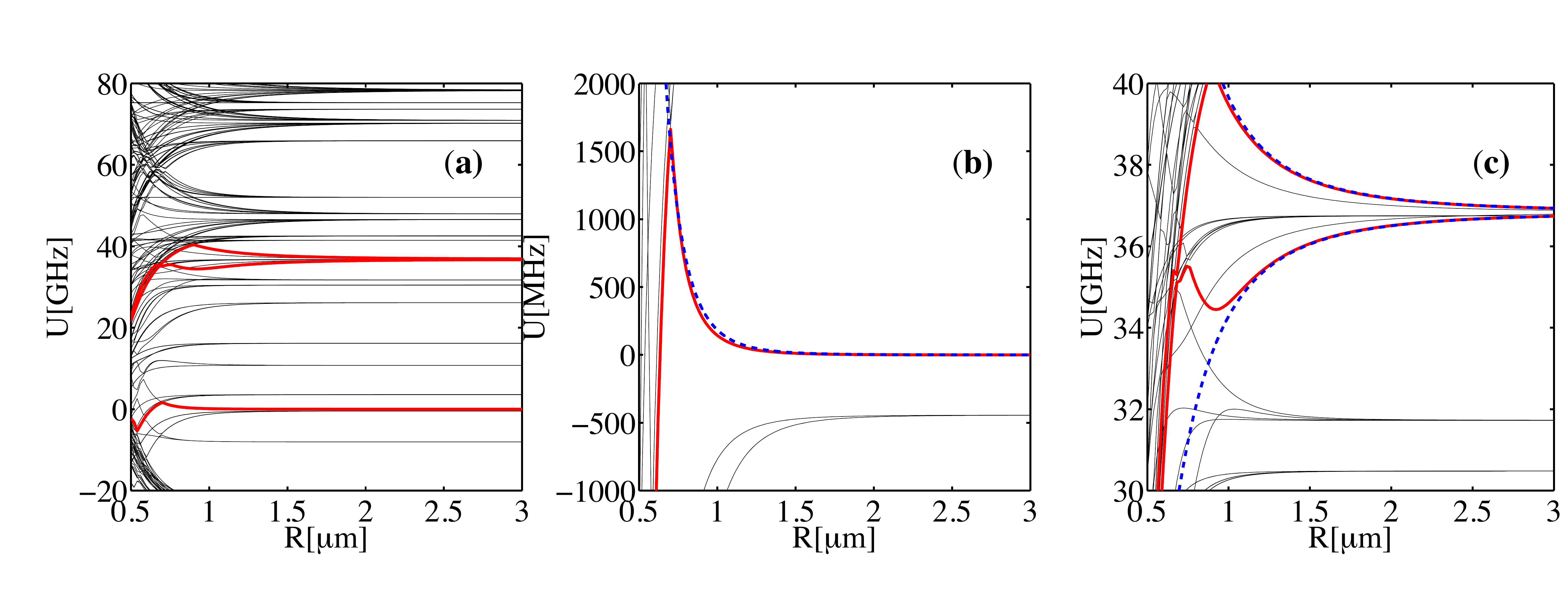} 
\caption{Rydberg interaction potentials for Lithium in the vicinity of the $\ket{40s0}\ket{40s0}$ asymptote (which is taken as the zero of energy), red curves are magnified in (b,c). (b) When zooming onto this potential at separations $R$ larger than $\sim0.8\mu$m, we recover the $U(R)=C_6/R^{6}$ dependence characteristic of vdW interactions (fit by the blue dashed line). (c) Resonant dipole-dipole interactions form potentials with asymptotic form $U_\pm(R)=\pm C_3/R^{3}$ (blue dashed) and states  $\ket{\varphi_\pm}=(\ket{sp} \pm\ket{ps})/\sqrt{2}$. 
\label{spaghetti_onset}}
\end{figure}
Approaching the end of this review, we now provide a brief guide to identify those regions in the vast parameter space for Rydberg physics, where either static- or flexible Rydberg aggregates as reviewed above can be practically realized. To this end, let us have a closer look at the origin and limitations of the simple model \bref{Hdd}, by considering the full picture of Rydberg-Rydberg atomic interactions. Most Rydberg potentials can be obtained by exact diagonalization of the Hamiltonian in a restricted Hilbert space with principal quantum numbers $\nu\in[\sub{\nu}{min}\dots \sub{\nu}{max}]$, angular momenta $l\in[0 \dots \sub{l}{max}]$ and the corresponding ranges of total angular momenta $j$ and their azimuthal quantum numbers $m_j$ \cite{fabian:soliton,walker:zeemandeg:pra,weber:rydint:tutorial}. The inter-atomic distance $R$ is varied and interactions can be added through the dipole-dipole interaction Hamiltonian \cite{book:gallagher,noordam:interactions,park:dipdipbroadening}.
 Recently, open source packages for this sort of calculation have become available \cite{weber:rydint:tutorial,pairinteraction:webpage,arc:package}.
 
  Alternatively, interactions can also be calculated using perturbation theory \cite{singer:VdWcoefficients}. The details listed are atomic species dependent; we will focus on Lithium atoms, because they are light and therefore of advantage to see motional effects, and on Rubidium atoms, because they represent the most common species in cold atom experiments.
 
 An exemplary set of molecular potentials is shown in \frefp{spaghetti_onset}{a}, in the vicinity of the $\ket{40s0}\ket{40s0}$ asymptote of Lithium. In this particular example, panel (c) shows that for inter-atomic separations $R>2\mu$m, and when a p-excitation is present in the systems, interactions are clearly dominated by resonant dipole-dipole interactions, involving just a pair of sp states. We can thus employ the model \bref{Hdd}. Since this model does not contain any interactions between two atoms in an $s$-state, we can augment it by a van-der-Waals term
\begin{equation}
\sub{\hat{H}}{vdW}=\sum_n E_n(\bv{R})\ket{\pi_n}\bra{\pi_n},
\\
E_k(\bv{R})\approx \sum_{\ell \ne n } \frac{C^{(sp)}_6}{R_{n\ell}^6}+\frac{1}{2}\sum_{j\ne n}\sum_{\ell \ne n,j } \frac{C^{(ss)}_6}{R_{j\ell}^6},
\label{HVdW}
\end{equation}
where $R_{n\ell}=|R_n-R_{\ell}|$ and $C^{(ab)}_6$ denotes the coefficient for the vdW interaction between one atom in the state $a$ and the other one in $b$.

This extension  also describes the short range repulsion (for $C_6>0$) shown in panel (b) down to $R>1\mu$m. At even shorter distance $R<\sub{R}{mix}$, we enter the highly irregular and dense region of the molecular spectrum shown in (a). Here simple models such as \bref{Hdd} and \bref{HVdW} break down entirely. Theory predictions presented here then no longer hold, and we thus anticipate that this region is best avoided. It might give rise to interesting experimental observations, but the most likely result of close encounters of Rydberg atoms is their ionization.

A rough estimate of the distances at which the spectra become dense is provided by he formula
\begin{equation}
\sub{R}{mix}(\nu)=2 \left( \frac{C_3(\nu)}{{\Delta}E_{pd}(\nu)}  \right)^{\frac{1}{3}}, 
\label{Rmix}
\end{equation}
where ${\Delta}E_{pd}(\nu)$ is the energy gap between the angular momentum $p$ and $d$ states at principal quantum number $\nu$, see  \aref{parameter_criteria} for further discussion. We empirically find the formula to work well for both species considered here, Li and Rb, based on spectra such as shown in \fref{spaghetti_onset}.

We are now in a position to determine for which parameters Rydberg aggregates described by the models summarized in this review can be formed, applying the following  criteria:
\begin{enumerate}
\item The effective state model should be valid, implying $d>\sub{R}{mix}$, where $d$ denotes any separation between Rydberg excited atoms in the system.
\item The dynamics of interest, static excitation transport or atomic motion, should take place within the life-time $\sub{\tau}{eff} =\tau_0/N$ of the system, where $\tau_0$ is the life-time of a single Rydberg state, and $N$ the number of Rydberg atoms in the aggregate. These effective  life-times become dependent on temperature $T$ by taking into account black-body redistribution of states \cite{beterov:BBR}.
\end{enumerate}

\begin{figure}[htb]
\centering
\includegraphics[width= 0.8\columnwidth]{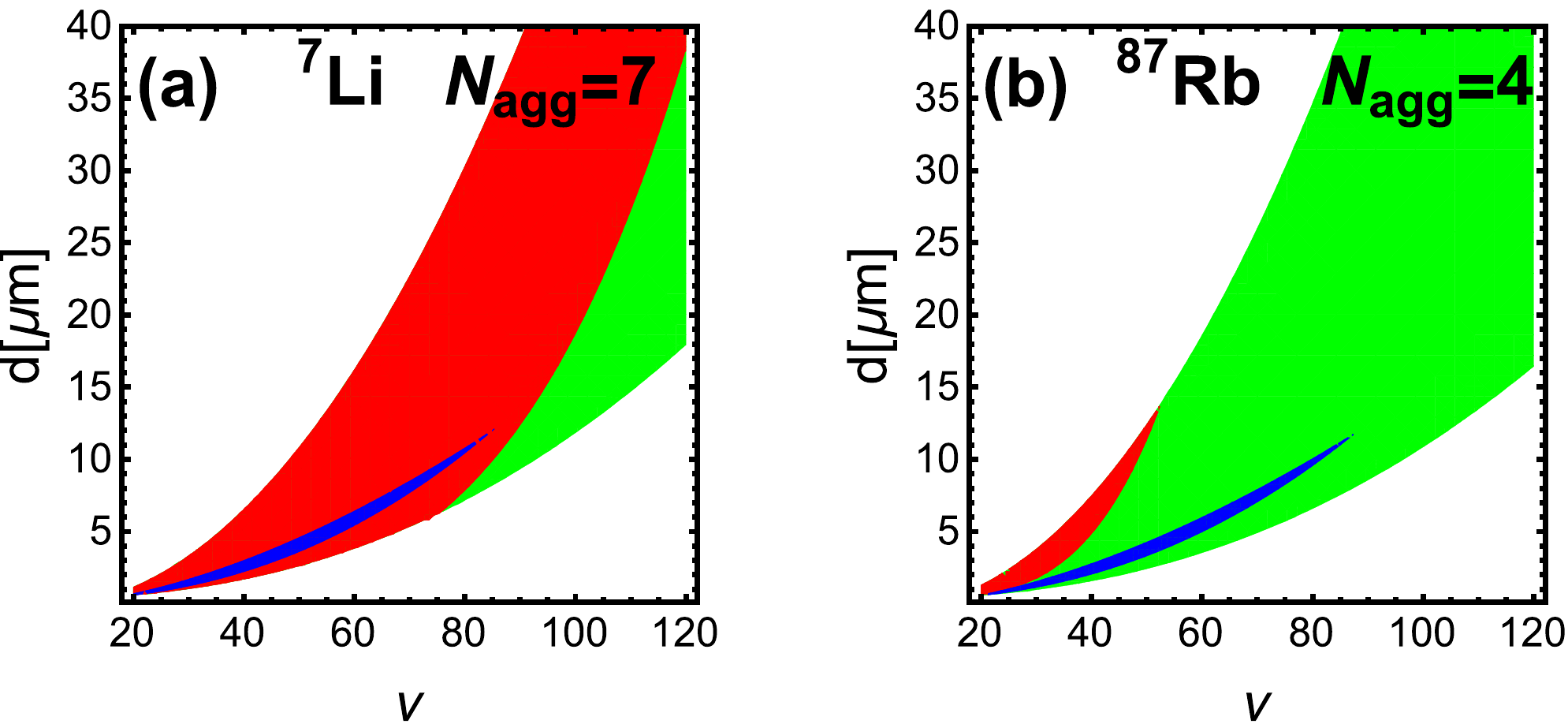} 
\caption{Parameter domains of static (green and red) versus flexible (blue) Rydberg aggregates. We take into account the reduction of effective life-times by black-body radiation at room temperature $T=300$K. Red shade indicates where static aggregates exist, however with atoms that would visibly accelerate during excitation transport. We consider the atomic species Li (a) and Rb (b), but importantly assume different aggregate sizes $\sub{N}{agg}$ for either as indicated.  White areas are excluded, either due to to short aggregate life-times (top) or too close proximities to avoid Rydberg state mixing (bottom). See the text and \aref{parameter_criteria} for the precise criteria used.
\label{parameter_ranges_T300}}
\end{figure}

Visualizing these conditions, we show in \fref{parameter_ranges_T300} the domain in the parameter plane $(\nu,d)$ where static- or flexible aggregates are viable at room temperature.
 We focus on the principal quantum number $\nu$ and the dominant inter-atomic spacing $d$, since these two parameters have most impact on the physics.
We consider aggregates with $\sub{N}{agg}$ atoms of the dominant experimental atomic species, $^{87}$Rb, or the lightest Alkali species with quantum defect $^{7}$Li. For simplicity we assumed a linear chain with dominant spacing $d$, and target dynamics such as in \fref{ballistic_transport} (static) or \fref{newtons_cradle} (flexible). The white region at the bottom is excluded due to $d<\sub{R}{mix}$. For the white region at the top, the overall effective life-time of the aggregate including black-body radiation (BBR) \cite{beterov:BBR} is too short for significant excitation transport (defined in more detail in \aref{parameter_criteria}) to take place.

Within these two boundaries we additionally distinguish: (i) The domain of flexible Rydberg aggregates (blue), defined in more detail in \aref{parameter_criteria}. (ii) The domain of purely static Rydberg aggregates (green). Here, for the duration of anticipated excitation transport, atoms would be trapped by their inertia and not significantly set into motion. (iii) Aggregates perturbed by atom acceleration (red). Here, atoms would significantly accelerate during the duration of excitation transport. We see that Lithium atoms can form larger flexible aggregates and be set into motion faster, all due to the much smaller atomic mass.

\begin{figure}[htb]
\centering
\includegraphics[width= 0.8\columnwidth]{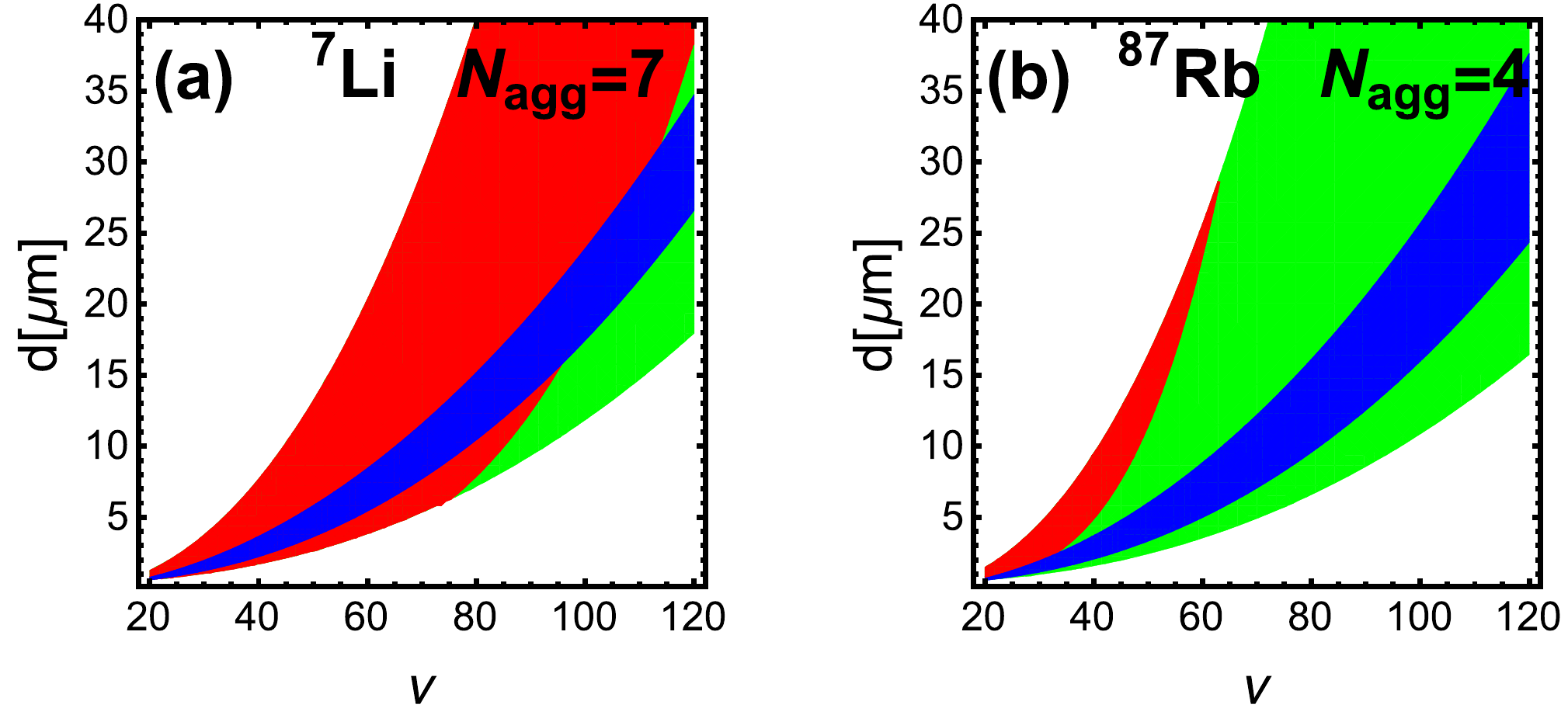} 
\caption{The same as \fref{parameter_ranges_T300} but assuming an extremely cold environment at $T=0$K.
\label{parameter_ranges_T0}}
\end{figure}

The domain for flexible Rydberg aggregates is strongly affected by the reduction of effective life-times due to BBR \cite{beterov:BBR}, hence we also show the corresponding domains when suppressing BBR at $T=0$K in \fref{parameter_ranges_T0}.
%

\section{Perspectives and outlook}\label{outlook}

We have seen that (flexible) Rydberg aggregates present a well defined, controllable and observable system in which to study the interplay of quantum coherent atomic motion and excitation.
Besides the fundamental interest in such a system, we see the following perspectives for their utilization: 

\ssection{Quantum simulations of nuclear dynamics in molecules} We have seen in \sref{conical_intersections} that excitonic Born-Oppenheimer surfaces exhibit conical intersections, one of the most fascinating features of molecular dynamics. In contrast to the intra-molecular situation, motional dynamics of Rydberg aggregates ought to be observable on scales relevant for the CI. It thus seems possible to furnish a quantum simulation platform that can either: (i)  replicate fundamental dynamical effects, such as vibrational relaxation downwards a CI funnel  \cite{Perun-Domcke-Abinitiostudies-2005}, and gain  new insight through a more direct connection between theory and observation; or (ii) attempt to mimick as closely as possible computationally known BO surfaces for a certain molecule and thus function as an analogue quantum computer to tackle the formidable computational challenge of multi-dimensional nuclear dynamics. Such a platform would then complement others, enabling the quantum simulation of \emph{electronic} dynamics in molecules with cold atoms \cite{Luehman_molorbquantsim_prx}.

\ssection{Quantum simulations of energy transport in biological systems} While this review has focused more strongly on the description of flexible Rydberg aggregates, we have listed several works that enable the controlled introduction of decoherence to static Rydberg aggregates in \sref{embedded}. Since decoherence, particularly through internal molecular vibrations, is a dominating feature of energy transport through dipole-dipole interactions on a molecular aggregate (such as a light-harvesting complex), this paves the way towards the utilization of these systems for photosynthetic energy transport simulations.

\section{Summary and conclusions}\label{summary}

We have reviewed  Rydberg aggregates, interacting assemblies of few Rydberg atoms that can exhibit excitation sharing and transport through mutual resonant dipole-dipole interactions. 
Our overview distinguished static aggregates, where atomic motion can be neglected during the times-scale of an experiment, and flexible aggregates, where atomic motion is by design an essential feature of the system. All results where placed into a broader context, considering the exploitation of these aggregates for quantum simulation applications, mainly by making contact with quantum chemistry or biology.  In this direction we have summarized a variety of theory proposals. For both, static- and flexible aggregates, we additionally gave an overview of  experimental techniques that are closest to the requirements for the implementation of those proposals.

It appears that the full theoretical potential of Rydberg aggregates in either variety discussed here is only just now becoming clear. Since experiments are now capable to provide all tools required for their controlled creation and observation, we expect a versatile and interdisciplinary quantum simulation platform in the near future.

\ack
We would like to thank C.~Ates, A.~Eisfeld, M. Genkin, K. Leonhardt, S.~M\"obius and H.~Zoubi for their part
in our joint publications featured in this review. Additionally, we thank Karsten Leonhardt for help with figures as well as a critical reading of the manuscript, and the authors of articles from which we reprint figures for their kind permission to do so.

\appendix

\section{Criteria for Rydberg aggregate parameter choices}
\label{parameter_criteria}
%
We have provided parameter space surveys in \sref{interactions}, clarifying under which conditions the main phenomena reviewed here, transport by excitation hopping with frozen atoms and adiabatic excitation transport with mobile atoms, are likely to be observable. To this end we cast a variety of physical requirements into mathematical criteria. 

\ssection{Validity of the essential state model} We have calculated interaction potentials exactly as in \fref{spaghetti_onset} for Li and Rb for a variety of principal quantum numbers $\nu$ in the range $20$ to $100$. The corresponding graphs permit an approximate visual identification of the respective $\sub{R}{mix}$ below which the spectra become highly chaotic. The manually determined  $\sub{R}{mix}(\nu)$ are well fit by the empirical formula \bref{Rmix}. In that formula we  obtain the dipole-dipole coefficient $C_3$ from a reference value $C^{(0)}_3$ using its  scaling $C_3 =  C^{(0)}_3\nu^4$, and similarly for ${\Delta}\sub{E}{pd}$  via  ${\Delta}\sub{E}{pd} = {\Delta}\sub{E}{pd}^{(0)} \nu^{-3}$. We have used the values $C^{(0)}_3=0.64$ a.u.~and ${\Delta}\sub{E}{pd}^{(0)}=0.05$ a.u.~for both species.

\ssection{Life-times} We have used the values tabulated in \cite{beterov:BBR}, with spline interpolation for intermediate $\nu$.

\ssection{Static aggregates} From \bref{Hdd} we can infer a transfer time (Rabi oscillation period) $\sub{T}{hop}=\pi d^3/C_3$ for an excitation to migrate from a given atom to the neighboring one, if the inter-atomic spacing is $d$. We have calculated the corresponding time for $\sub{N}{hops}$ such transfers, given by $\sub{T}{trans} =\sub{N}{hops} \sub{T}{hop}$, imagining migration along an entire aggregate. Here $\sub{N}{hops}$ can be quite freely chosen, we have opted for $\sub{N}{hops}=100$ to include possibly sophisticated transport. We finally demand $\sub{T}{trans}$ to be short compared to the system life-time.

\ssection{Perturbing acceleration} By solving the differential equation of motion for a single Rydberg dimer, which is accelerated by repulsive dipole-dipole interactions from rest, we can infer a characteristic time-scale  scale for acceleration $\sub{T}{acc}=\sqrt{\frac{R_0^5 M}{6 C_3}}$. The mass of the atoms is $M$ and their initial separation $R_0$. When $\sub{T}{acc}$ becomes less than $\sub{T}{trans}$, we color static aggregates red in \fref{parameter_ranges_T300} and \fref{parameter_ranges_T0}.

\ssection{Flexible aggregates} For flexible aggregates, we assume an equidistant chain with spacing $d$, but the existence of a dislocation on the first two atoms with spacing of only $\sub{d}{ini}=2 d/3$ to initiate directed motion, as in \fref{newtons_cradle}. Hence, $\sub{d}{ini}>\sub{R}{mix}$ must be fullfilled, a tighter constraint than $d>\sub{R}{mix}$. This results in the shift towards larger $d$ of the minimal $d$ for flexible aggregates compared to static aggregates in \fref{parameter_ranges_T300} and \fref{parameter_ranges_T0}. The velocity  $v$ is then determined from the initial potential energy $\sub{E}{ini}= C_3/\sub{d}{ini}^3$ via energy conservation, and the duration of interesting motional dynamics is set to $\sub{T}{mov} = \sub{T}{acc} + (\sub{N}{agg}-2)d/v$, approximating the time for a joint excitation-dislocation pulse such as in \fref{newtons_cradle} to traverse the entire aggregate. Again, we require $\sub{T}{mov}$ to be short compared to the system life-time.

\section*{References}
\bibliographystyle{sebastian_v3}
\bibliography{rydberg_aggregates}

\begin{thebibliography}{100}
\bibitem{frenkel:footnote}
Some confusion can arise from the varied use of (Frenkel) exciton in the
  condensed matter, quantum chemistry and molecular aggregate community. The
  former implies an electron-hole pair, with hole localised on a crystal site.
  For a single molecule, and excited state corresponds to a Frenkel exciton
  with hole in the LUMO and electron in the HOMO orbital. Finally for molecular
  aggregates, Frenkel excitons imply the delocalization of the latter state
  over the aggregate due to interactions. Our definition is based on the
  analogy to the latter.
\end{thebibliography}
\end{document}